\documentclass[aps,nofootinbib,showkeys,a4paper]{revtex4}
\usepackage{graphicx}

\usepackage{supertabular}

\usepackage{psfrag}

\def\p{\partial}
\def\a{\alpha}

\def\d{\delta}
\def\D{\Delta}
\def\e{\epsilon}
\def\ee{\varepsilon}

\def\l{\lambda}

\def\s{\sigma}
\def\t{\tau}
\def\ra{\rightarrow}

\def\O{{\cal O}}
\def\T{{\cal T}}

\def\Mfunction#1{\mathop{\rm #1}\nolimits}

\begin{document}
\title{Linear approach to the orbiting spacecraft thermal problem} 
\author{Jos\'e~Gaite and Germ\'an Fern\'andez-Rico}
\affiliation{
IDR, ETSI Aeron\'auticos, Universidad Polit\'ecnica de Madrid,
Pza.\ Cardenal Cisneros 3, E-28040 Madrid, Spain} 
\date{January~18, 2012}

\begin{abstract}
We develop a linear method for solving the nonlinear differential equations of
a lumped-parameter thermal model of a spacecraft moving in a closed orbit.
Our method, based on perturbation theory, is compared with heuristic
linearizations of the same equations. The essential feature of the linear
approach is that it provides a decomposition in thermal modes, like the
decomposition of mechanical vibrations in normal modes.  The stationary
periodic solution of the linear equations can be alternately expressed as an
explicit integral or as a Fourier series.  We apply our method to a minimal
thermal model of a satellite with ten isothermal parts (nodes) and we compare
the method with direct numerical integration of the nonlinear equations.  We
briefly study the computational complexity of our method for general thermal
models of orbiting spacecraft and conclude that it is certainly useful for
reduced models and conceptual design but it can also be more efficient than
the direct integration of the equations for large models.  The results of the
Fourier series computations for the ten-node satellite model show that the
periodic solution at the second perturbative order is sufficiently accurate.
\end{abstract}


\maketitle

\subsection*{Nomenclature}

\begin{supertabular}{lcl}
$A_i$ & = & outward-facing area of \emph{i}th-node, m$^{2}$\\
$C$ & = & thermal capacitance diagonal matrix, J/K\\
$C_{i}$ & = & thermal capacitance of \emph{i}th-node, J/K\\
$e_a$ & = & eigenvector of $C^{1/2}JC^{-1/2}$, K\\
$e_{aj}$ & = & \emph{j}th-component of eigenvector $e_a$, K\\

$F(t)$ & = & driving function vector, K/s\\

$\hat{F}(m)$ & = & $m$th Fourier coefficient of $F(t)$, K/s\\

$\hat{F}^*$ & = & complex conjugate of $\hat{F}$, K/s\\

$F_a(t)$ & = & driving function for $a$th-mode, K/s\\

$\dot{F}_a(t)$ & = & time derivative of $F_a(t)$, K/s$^2$\\

$F_i(t)$ & = &  driving function for $i$th-node, K/s\\

$\hat{F}_i(m)$ & = & $m$th Fourier coefficient of $F_i(t)$, K/s\\

$G(t)$ & = & driving function for second order temperature vector, K/s\\

$J$ & = & Jacobian matrix, s$^{-1}$\\

$J_{ij}$ & = & Jacobian matrix \emph{ij}-element, s$^{-1}$\\

$k$ & = & number of steps in numerical integration of ODEs\\

$K$ & = & conduction coupling matrix, W/K\\

$K^R$ & = & matrix of conductances from linearized radiation coupling terms, W/K\\

$K_{i}^{R}$ & = & $i$th-node conductance from linearized environment-radiation
  terms, W/K\\

$K_{ij}$ & = & conduction coupling matrix \emph{ij}-element, W/K\\

$K_{ij}^R$ & = & \emph{ij}-element of conductance matrix from linearized
  radiation terms, W/K\\ 

$n$ & = & number of samples for discrete Fourier transform\\

$N$ & = & number of nodes\\

$P$ & = & eigenvector matrix for \emph{J}, K\\

$P_{ia}$ & = & \emph{i}th-component of $J$-eigenvector for $a$th-mode, K\\

$q$ & = & auxiliary heat input vector, K/s\\

$\dot{Q}_i(t)$ & = & heat input to \emph{i}th-node, W\\

$\langle\dot{Q}_i\rangle$ & = & mean value of $\dot{Q}_i(t)$ over period $\cal{T}$, W\\

$R_i$ & = & \emph{i}th-node coefficient of radiation to the environment, W/K$^4$\\

$R_{ij}$ & = & radiation coupling matrix, W/K$^{4}$\\

$t$ & = & time, s (min in figures)\\

$T$ & = & temperature vector, K\\

$\Delta T$ & = & vector of ``errors'' w.r.t. ESATAN$^\mathrm{TM}$ solution, K\\

$\widetilde{T}$ & = & steady-state temperature vector, K\\

$T_{i}$ & = & temperature of \emph{i-}node, K\\

$T_{0}$ & = & cosmic microwave background radiation temperature, K\\

$T_{(n)j}$ & = & \emph{n}th order term in \emph{j}th-node temperature expansion, K\\

$T^{\infty}$ & = & stationary solution temperature vector, K\\

$T^{\infty}_{(n)}$ & = & \emph{n}th order of stationary solution temperature vector, K\\

$\dot{T}_i$ & = & time derivative of \emph{i}th-node temperature, K/s\\

$\widetilde{T}_i$ & = & steady-state temperature of \emph{i}th-node, K\\

$\cal{T}$ & = & orbital period, s \\

$U$ & = & matrix of independent solutions to homogeneous equations, K\\

$\alpha_s$ & = & solar absorptivity\\

$\delta_{ab}$ & = & Kronecker's delta\\

$\delta A$ & = & antisymmetric part of $C^{1/2}JC^{-1/2}$, s$^{-1}$\\
$\delta A_{ij}$ & = & \emph{ij}-element of $\delta A$, s$^{-1}$\\

$\varepsilon_i$ & = & infrared emissivity of \emph{i}th-node outward-facing surface\\

$\epsilon$ & = & perturbation parameter\\

$\lambda_a$ & = & eigenvalue of \emph{J} for $a$th mode, s$^{-1}$\\

$\delta \lambda_a$ & = & perturbation of \emph{J}-eigenvalue for $a$th mode,
  s$^{-1}$\\

$\sigma$ & = & Stefan-Boltzmann constant, 
$5.67\times10^{-8}$ W\hspace{2pt}m$^{-2}\,$K$^{-4}$\\
\end{supertabular}

\section{Introduction}
\label{intro}

The thermal control of a spacecraft ensures that the temperatures of its
various parts are kept within their appropriate ranges
\cite{Kreith,therm-control,therm-control_2,therm-control_3}. 
The simulation and prediction of temperatures in a spacecraft during a mission
are usually carried out by commercial software packages.  These software
packages employ ``lumped parameter'' models that describe the spacecraft as a
discrete network of nodes, with one energy-balance equation per node. The
equations for the thermal state evolution are coupled nonlinear first-order
differential equations, which can be integrated numerically.  Given the
thermal parameters of the model and its initial thermal state, the numerical
integration of the differential equations yields the solution of the
problem, namely, the evolution of the node temperatures. However,
a detailed model with many nodes is difficult to handle, and its integration
for a sufficiently long time of evolution can take considerable computer time
and resources. Therefore, it is very useful to study simplified models and
approximate methods of integrating the differential equations.

Many spacecraft missions, in particular, satellite missions, consist of an
initial transient part and then a stationary part, in which the spacecraft
just goes around a closed orbit, in which the heat inputs are periodic. These
periodic heat inputs are expected to induce periodic temperature variations,
with a maximum and a minimum temperature in each orbit.  This suggests a
conservative approach that consists in computing only the temperatures for the
{\em hot} and {\em cold cases} of the given orbit, defining them as the two
steady cases with the maximum and minimum heat loads, respectively.
Naturally, the real temperature variations in the orbit are smaller, because
there is not enough time for the hot and cold cases to establish themselves.
In fact, the temperature variations can be considerably smaller, to such a
degree that it is necessary to integrate the differential equations, at least
approximately.

The differential equations for energy balance are nonlinear due to the
presence of radiation couplings, which follow the Stefan-Boltzmann quartic
law.  A common approach to these equations involves a linearization of the
radiation terms that approximate them by heat conduction terms
\cite{anal-sat,therm-control_2,anal-sat_2,IDR}.  This approach transforms the
nonlinear equations into standard linear heat conduction equations.  But this
approach has not been sufficiently justified, is of a heuristic nature and
does not constitute a systematic approximation.

In fact, nonlinear equations are very different from linear equations and, in
particular, a periodic driving may not induce periodic solutions but much more
complex solutions, namely, {\em chaotic} solutions. Therefore, we have carried
out in preceding papers a full nonlinear analysis of spacecraft thermal models
\cite{NoDy,NoDy1}.  The conclusion of the analysis is that the complexities of
nonlinear dynamics, such as multiple equilibria and chaos, do not appear in
these models.  While the existence of only one equilibrium state can be proved
in general, the absence of chaos under driving by variable external heat loads
can only be proved for a limited range of magnitudes of the driving
loads. This range presumably includes the magnitudes involved in typical
spacecraft orbits.  The proofs in Refs.~\citenum{NoDy} and \citenum{NoDy1} are
constructive and are based on a perturbation method that is expected to be
sound when the {\em linear} equations corresponding to the first perturbative
order constitute a good approximation of the nonlinear equations.  This
implies that the fully nonlinear solution describes a weakly nonlinear
oscillator.  Since the perturbative approximation is mathematically rigorous
and systematic, it is worthwhile to study in detail the scope of the
perturbative linear equations and, furthermore, to compare them with previous
linear approaches of a heuristic nature.

The main purpose of this paper is to study the linear method of predicting the
thermal behavior of spacecraft in stationary orbits (Sect.\ \ref{linear} and
\ref{Fourier-anal}) and to test it on a minimally realistic thermal model of a
satellite in a circular orbit.  Since the general one and two-node models
analyzed in Refs.~\citenum{NoDy} and \citenum{NoDy1}, respectively, are too
simple, we define in this paper a ten-node thermal model of a small
Moon-orbiting satellite (Sect.~\ref{ten-node}). This model is simple enough to
allow us to explicitly show all the quantities involved (thermal couplings and
capacities, heat inputs, etc.) and it is sufficient for illustrating the main
features of the linear approach.  As realistic thermal models have many more
nodes, we consider in Sect.~\ref{scale} the important issue of scalability of
the method and, hence, its practical applications.  Computational aspects of
the steady-state problem have been studied by Krishnaprakas \cite{Kris1,Kris2}
and by Milman and Petrick \cite{MiPe}, while computational aspects of the
direct integration of the nonlinear equations for the unsteady problem have
been studied by Krishnaprakas \cite{Kris3}.  Here we focus on the linear
equations for the stationary but unsteady case and survey its computational
aspects.

A note on notation: In the equations that contain matrix or vector
quantities, sometimes we use component notation (with indices) while
other times we use compact matrix notation (without indices), according
to the nature of the equations. 

\section{Linearization of the heat-balance equations}
\label{linear}

A lumped-parameter thermal model of a continuous system consists of a discrete
network of isothermal regions ({\em nodes}) that represent a partition of the
total thermal capacitance and that are linked by thermal conduction and
radiation couplings
\cite{Kreith,therm-control,therm-control_2,therm-control_3,anal-sat}.  This
discretization reduces the integro-differential heat-transfer equations to a
set of energy-balance ODEs, one per node, which control the evolution of the
nodes' temperatures \cite{anal-sat}:
\begin{eqnarray}
\label{ODE}
C_i \, \dot{T_i} = \dot{Q}_i(t) - 
\sum_{j=1}^N \left[K_{ij} (T_i - T_j) + R_{ij} (T_{i}^4 - T_{j}^4)
\right] - R_i\,({T_i}^4- T_{0}^4),
\quad i= 1,\ldots,N,
\end{eqnarray}
where $N$ is the number of nodes and $\dot{Q}_i(t)$ contains the total heat
input to the $i$th-node from external radiation and from internal energy
dissipation (if there is any).  The conduction and radiation coupling matrices
are denoted by $K$ and $R$, respectively; they are symmetric ($K_{ij} =
K_{ji}$ and $R_{ij} = R_{ji}$) and $K_{ii}=R_{ii}=0$; so there are $N(N-1)$
independent coupling coefficients altogether, but many vanish, usually.  The
temperature $T_0 \simeq 3\!$~K is the temperature of the environment, namely,
the cosmic microwave background radiation.  The $i$th-node coefficient of
radiation to the environment is given by $R_i=A_i\ee_i\s$, where $A_i$ denotes
the outward facing area, $\ee_i$ its (infrared) emissivity, and $\s$ is the
Stefan-Boltzmann constant.  The constant term $R_i T_0^4$ can be included in
$\dot{Q}_i(t)$ or ignored altogether, if each ${T_i} \gg T_0$.
Equations~(\ref{ODE}) coincide with the ones implemented in commercial
software packages, for example, ESATAN$^\mathrm{TM}$ \cite{ESATAN}.

There is no systematic procedure for finding the analytical solution of a system
of nonlinear differential equations, except in some particularly simple
cases. Of course, nonlinear systems can always be integrated numerically with
finite difference schemes. Methods of this kind are employed in commercial
software packages.  When a nonlinear system can be approximated by a linear
system and, hence, an approximate analytic solution can be found, this
solution constitutes a valuable tool.  Actually, one can always resort to some
kind of perturbation method to linearize a nonlinear system.  Therefore, we
now study the rigorous linearization of Eqs.~(\ref{ODE}) based on a suitable
perturbation method, and we also describe, for the sake of a comparison, a
heuristic linearization, which actually is best understood in light of the
results of the perturbation method.

\subsection{Perturbative linearization}
\label{pert-eq}

If we assume that the heat inputs $\dot{Q_i}(t)$ in the energy-balance
Eqs.~(\ref{ODE}) are periodic, namely, that there is a time interval $\T$ such
that $\dot{Q_i}(t+\T) = \dot{Q_i}(t)$, then it seems sensible to study first
the effect of the mean heat inputs in a period. This {\em averaging} method,
introduced in Refs.~\citenum{NoDy} and~\citenum{NoDy1}, relies on the fact
that the {\em autonomous} nonlinear system of ODEs for constant $\dot{Q_i}$
can be thoroughly analyzed with analytical and numerical methods. For example,
it is possible to determine that there is a unique steady thermal state and
that it is (locally) stable \cite{MiPe,NoDy1}.  The actual values of the
steady temperatures can be found efficiently with various numerical methods
\cite{Kris1,Kris2,MiPe}.  Furthermore, the eigenvalues and eigenvectors of the
Jacobian matrix of the nonlinear system of ODEs provides us with useful
information about the dynamics, in particular, about the approach to
steady-state: the eigenvectors represent independent thermal modes and the
eigenvalues represent their relaxation times \cite{NoDy1}.

Once the averaged equations are solved, the variation of the heat inputs can
be considered as a driving of the averaged solutions.  Thus, we can define the
driving function
$$
F_i(t) = \frac{\dot{Q_i}(t) - \langle{\dot{Q_i}}\rangle}{C_i}\,, 
\quad i=1, \ldots, N,
$$
where $\langle{\dot{Q_i}}\rangle$ denotes the mean value of $\dot{Q_i}(t)$
over the period of oscillation. A weak driving function must not produce a
notable deviation from the averaged dynamics. In particular, the long-term
thermal state of an orbiting spacecraft must oscillate about the corresponding
steady-state.  To embody this idea, we introduce a formal perturbation
parameter $\e$, to be set to the value of unity at the end, and write
Eqs.~(\ref{ODE}) as
\begin{eqnarray}
\label{eODE}
\dot{T}_i = \e \, F_i(t) + \frac{\langle{\dot{Q_i}}\rangle}{C_i} 
- 
\sum_{j=1}^N \left[\frac{K_{ij}}{C_i} 
(T_i - T_j) + \frac{R_{ij}}{C_i}  (T_{i}^4 - T_{j}^4)
\right] - \frac{R_{i}}{C_i} \,{T_i}^4,
\quad i= 1,\ldots,N,
\end{eqnarray}
Then, we assume an expansion of the form
\begin{equation}
\label{Tseries}
T_j(t) = \sum_{n=0}^{\infty} \e^n\, T_{(n)j}(t)\,.
\end{equation}
When we substitute this expansion into Eqs.~(\ref{eODE}),
we obtain for the zeroth order of $\e$
\begin{equation}
\label{bODE-pert}
\dot{T}_{(0)i} =  \frac{\langle{\dot{Q_i}}\rangle}{C_i}
- \sum_{j=1}^N \left[\frac{K_{ij}}{C_i} (T_{(0)i} - T_{(0)j}) +
\frac{R_{ij}}{C_i} (T_{(0)i}^4 - T_{(0)j}^4) \right] - \frac{R_{i}}{C_i}
\,T_{(0)i}^4\,,
\quad i= 1,\ldots,N,
\end{equation}
that is to say, the averaged equations. The initial conditions for these
equations are the same as for the unaveraged equations. 

For the first order in $\e$, we obtain the following system of linear
equations:
\begin{equation}
\label{linODE-pert}
\dot{T}_{(1)i} =  \sum_{j=1}^N J_{ij}(t) \,{T}_{(1)j} + F_i(t)\,,
\quad i=1, \ldots, N.
\end{equation}
Here, $J_{ij}(t)$ is the Jacobian matrix
$$
J_{ij}(t) = \left.\frac{\p}{\p T_j}\dot{T}_i(T) \right|_{T=T_{(0)}(t)},
$$
where $T_{(0)}(t)$ is the solution of the zeroth order equation.
Equations~(\ref{linODE-pert}) are to be solved with the initial condition
$T_{(1)}(0)=0$.

The elements of the Jacobian matrix at a generic point in the temperature
space are calculated to be: 
{\setlength\arraycolsep{2pt}
\begin{eqnarray}
\label{Jij}
J_{ij} &=&  C_i^{-1} \left(K_{ij} + 4 R_{ij} T_{j}^3\right),
\quad \mathrm{if}\;i \neq j,\\
\label{Jii}
J_{ii} &=& C_i^{-1} \left[-\sum_{k=1}^N \left(K_{ik} + 4 R_{ik} T_{i}^3
\right) - 4 R_i\,{T_i}^3\right].
\end{eqnarray}
}%
This matrix has interesting properties. First of all, it has negative diagonal
and nonnegative off-diagonal elements. In other words, $-J$ is a $Z$-matrix
\cite{Ber-Plem}. Furthermore, it fulfills a semipositivity condition that
qualifies it as a nonsingular $M$-matrix \cite{NoDy1}. Since the eigenvalues
of an $M$-matrix have positive real parts, the opposite holds for $J$, namely,
its eigenvalues have negative real parts.  One more interesting property of
$-J$, related to semipositivity, is that it possesses a form of {\em diagonal
  dominance}: it is similar to a diagonally dominant matrix and the similarity
is given by a positive diagonal matrix. Naturally, this property is shared by
$J$.  These properties are useful to prove some desirable properties of the
solutions of Eqs.~(\ref{linODE-pert}).

The chief property of $J$ is that $-J$ is a nonsingular $M$-matrix. In
particular, it implies that $-J^{-1}$ is non-negative and, therefore, that the
Perron-Frobenius theory is applicable to it \cite{Ber-Plem}. The relevant
results to be applied are: (i) Perron's theorem, which states that a {\em
  strictly} positive matrix has a unique real and positive eigenvalue with a
positive eigenvector and that this eigenvalue has maximal modulus among all
the eigenvalues; (ii) a second theorem, stating that if a $Z$-matrix that is a
nonsingular $M$-matrix is also ``irreducible'', then its inverse is strictly
positive.  The irreducibility of $J$ follows from the symmetry of the matrices
$K_{ij}$ and $R_{ij}$ \cite{NoDy1}.  As the positive (Perron) eigenvector of
$-J^{-1}$ is the eigenvector of $J$ that corresponds to its smallest magnitude
eigenvalue, it defines the slowest relaxation mode (for a given set of
temperatures).  Therefore, in the evolution of temperatures given by
Eqs.~(\ref{bODE-pert}), steady-state is eventually approached from the
zone corresponding to simultaneous temperature increments (or decrements).

The matrix $J(t)$ in Eqs.~(\ref{linODE-pert}) is obtained by substituting
$T_{(0)j}(t)$ for $T_{j}$ in Eqs.~(\ref{Jij}) and (\ref{Jii}).  Then, the
nonhomogeneous linear system with variable coefficients,
Eqs.~(\ref{linODE-pert}), can be solved by variation of parameters
\cite{NoDy1}, yielding the expression:
\begin{equation}
T_{(1)}(t) = U(t) \int_0^t U(\t)^{-1}\cdot F(\t) \,d\t,
\label{T_1}
\end{equation}
where $U(t)$ is a matrix formed by columns that are linearly independent
solutions of the corresponding homogeneous equation, with the condition that
$U(0) = I$ (the identity matrix). The difficulty in applying this formula lies
in computing $U(t)$, that is, in computing the solutions of the homogeneous
equation. Moreover, this computation demands the previous computation of the
solution for $T_{(0)}(t)$.

Since we are only interested in the stationary solutions of the heat-balance
equations rather than in transient thermal states, it is possible to find an
expression of these solutions that is more manageable than
Eq.~(\ref{T_1}). The transient thermal state relaxes exponentially to the
stationary solution, which is a {\em limit cycle} of the nonlinear equations,
technically speaking \cite{NoDy,NoDy1}.  Therefore, the stationary solution is
given by the solution of Eqs.~(\ref{linODE-pert}) with the {\em constant}
Jacobian matrix calculated at the steady-state temperatures, which we name
$\widetilde{T}_{i}, \; i= 1,\ldots,N$.%
\footnote{The solution can also be derived as the limit of 
Eq.~(\ref{T_1}) in which $U(t) = \exp(J t)$.}
This solution is simply \cite{NoDy1}:
\begin{equation}
{T}_{(1)}(t) = 
\int_0^t  \exp\left[\t J \right] \cdot F(t-\t) \,d\t,
\label{T_1_Jcst}
\end{equation}
with $J$ calculated at the point $\widetilde{T}$.  Furthermore, the periodic
stationary solution is obtained by extending the upper integration limit from
$t$ to infinity:
\begin{equation}
{T}_{(1)}^\infty(t) = \int_0^\infty \exp\left[\t J \right] \cdot F(t-\t)
\,d\t.
\label{T_1_lim}
\end{equation}
This function is indeed periodic, unlike the one defined by
Eq.~(\ref{T_1_Jcst}), so it is determined by its values for $t \in [0,\T]$.
Note that $\langle {T}_{(1)}^\infty(t) \rangle = 0$.  For numerical
computations, it can be convenient to express the integral from 0 to $\infty$
as an integral from 0 to $\T$, taking advantage of the periodicity as follows:
{\setlength\arraycolsep{2pt}
\begin{eqnarray*}
&&\int_0^\infty \exp\left[\t J \right] \cdot F(t-\t)\,d\t =
\sum_{n=0}^\infty \int_{n\T}^{(n+1)\T} \exp\left[\t J \right] \cdot
F(t-\t)\,d\t
=\\
&&\sum_{n=0}^\infty \exp(n\T J) \int_0^{\T} \exp\left[\t J \right] \cdot
F(t-\t)\,d\t = 
[I-\exp(\T J)]^{-1} \int_0^{\T} \exp\left[\t J \right] \cdot
F(t-\t)\,d\t
\end{eqnarray*}
}%
(the series converges because the eigenvalues of $J$ have negative real
parts).  In the last integral, the argument of $F$ can be transferred to the
interval $[0,\T]$:
$$
\int_0^{\T} \exp\left[\t J \right] \cdot F(t-\t)\,d\t =
\int_0^{t} \exp\left[\t J \right] \cdot F(t-\t)\,d\t +
\int_t^{\T} \exp\left[\t J \right] \cdot F(t-\t+\T)\,d\t,
$$
where $t \in [0,\T]$. Note that the one-period shift in the argument of the
last  $F$ is necessary for the argument to be in $[0,\T]$.

Some remarks are in order. First of all, we have assumed that there is one
asymptotic periodic solution of the nonlinear Eqs.~(\ref{eODE}) and only
one (a unique limit cycle). Equivalently, we have assumed that the
perturbation series converges. This assumption holds in an interval of the
amplitude of heat input-variations $F$ \cite{NoDy1}. Besides, for the
integrals in Eq.~(\ref{T_1_lim}) and the following equations to make sense, it
is required that $\exp\left[\t J \right] \ra 0$ as $\t\ra\infty$. This is
guaranteed, because the eigenvalues of $J$ have negative real parts, as is
necessary for the steady-state to be stable.  In fact, the eigenvalues are
expected to be negative real numbers and $J$ is expected to be diagonalizable
but both properties are not rigorously proven \cite{NoDy1} (however, see
Sect.~\ref{heur}).

If $J$ is diagonalizable, that is to say, there is a real matrix $P$ such that
$P^{-1} J P$ is diagonal, then the calculation of the integrals is best
carried out on the eigenvector basis, given by the matrix $P$.  Using this
basis, Eq.~(\ref{T_1_lim}) is expressed as
\begin{equation}
\left[{T}_{(1)}^\infty\right]_i(t) = 
\sum_{a=1}^N P_{ia} 
\int_0^\infty \exp\left[\t \l_a \right] \sum_{j=1}^N P^{-1}_{aj}
F_j(t-\t)\,d\t, 
\quad i= 1,\ldots,N,
\label{T_1_eigenvec}
\end{equation}
where the first sum runs over the eigenvectors and their corresponding
eigenvalues $\l_a$.  Expression (\ref{T_1_eigenvec}) allows us to compare the
contribution of the different thermal modes.  In particular, for the fast
modes, such that $|\l_a|$ is large, we can use Watson's lemma
\cite{pert-methods} to derive the asymptotic expansion:
$$
\int_0^\infty \exp\left[\t \l_a \right] F_a(t-\t)\,d\t
= \frac{F_a(t)}{-\l_a} - \frac{\dot{F}_a(t)}{\l_a^2} 
+ \Mfunction{O}\left(\frac{1}{\l_a^3} \right), 
$$
where $F_a = \sum_{j} P^{-1}_{aj} F_j$.  When $|\l_a|$ is large, the first
term suffices (unless $\dot{F}_a(t)$ is also large, for some reason); and the first
term is small, unless $F_a(t)$ is large. In essence, if the fast modes are
not driven strongly, they can be neglected in the sum over $a$ in
Eq.~(\ref{T_1_eigenvec}).

\subsubsection{Second order perturbative equation}
\label{second-order}

For second order in $\e$, a straightforward calculation \cite{NoDy1} yields
the following linear equation: 
\begin{equation}
\label{linODE-pert2}
\dot{T}_{(2)} =   J(t) \cdot {T}_{(2)} + G(t)\,,
\end{equation}
where $J(t)$ is the same Jacobian matrix that appears in the first-order
Eq.~(\ref{linODE-pert}) and
\begin{equation}
\label{tF}
G_i = 
\sum_{j=1}^N\frac{6 \,R_{ij}}{C_i}\,T_{(0)j}^2\,T_{(1)j}^2 -
\frac{6}{C_i}\left(\sum_{j=1}^N R_{ij}+ R_{i}\right) T_{(0)i}^2\,T_{(1)i}^2
\,,
\quad i= 1,\ldots,N. 
\end{equation}
The initial condition for Eq.~(\ref{linODE-pert2}) is $T_{(2)}(0)=0$, as for
Eqs.~(\ref{linODE-pert}).  Therefore, the first-order and second-order
equations have identical solutions in terms of their respective driving terms,
although $G$, Eqs.~(\ref{tF}), is a known function of $t$ only when the lower
order equations have been solved.  The integral expression,
Eqs.~(\ref{T_1_lim}), of the stationary solution ${T}^\infty_{(1)}(t)$ is also
valid for ${T}^\infty_{(2)}(t)$, after replacing $F$ with $G$ and using in
Eq.~(\ref{tF}) the stationary values $T_{(0)}(t)=\widetilde{T}$ and
$T_{(1)}(t)={T}^\infty_{(1)}(t)$ (which make $G$ periodic).

It is possible to carry on the perturbation method to higher orders, and it
always amounts to solving the same linear equation with increasingly
complicated driving terms that involve the solutions of the lower order
equations.  The example of Sect.~\ref{ten-node} shows that, in a typical case,
${T}^\infty_{(2)}(t)$ is a small correction to ${T}^\infty_{(1)}(t)$, and
further corrections are not necessary. This confirms that the perturbation
method is reliable for a realistic case.

\subsection{Heuristic linearization}
\label{heur}

A linearization procedure frequently used in problems of radiation heat
transfer \cite{therm-control_2,anal-sat_2,IDR} consists of using the algebraic
identity
$$
T_{i}^4 - T_{j}^4 = (T_{i} + T_{j})(T_{i}^2 + T_{j}^2)(T_{i} - T_{j})
$$
to define an effective conductance for the radiation coupling
between nodes $i$ and $j$. The equation
$$
R_{ij} (T_{i}^4 - T_{j}^4) = K^R_{ij} (T_i - T_j),
$$
defines the effective conductance
$$K^R_{ij} = R_{ij} (T_{i} + T_{j})(T_{i}^2 + T_{j}^2)
$$ 
for specified values of the node temperatures $T_{i}$ and $T_{j}$.  For an
orbiting spacecraft, the natural base values of the node temperatures are the
ones that correspond to the steady-state solution of the averaged equations,
namely, $\widetilde{T}_{i}, \; i= 1,\ldots,N.$ In the special case of
radiation to the environment, $R_i T_{i}^4$ can be replaced with
linear terms $K_{i}^R\, T_{i}$ such that $K_{i}^R = 4 \widetilde{T}_{i}^3
R_i,$ for $i= 1,\ldots,N.$

The resulting linear equations are:
\begin{eqnarray}
\label{linODE}
C_i \, \dot{T_i} = \dot{Q}_i(t) - 
\sum_{j=1}^N \left(K_{ij} + K^R_{ij} \right)(T_i - T_j)
 - K^R_i\,{T_i}\,,
\quad i= 1,\ldots,N,
\end{eqnarray}
These equations have only conduction couplings, so they are a discretization
of the partial differential equations of heat conduction.  As a linear system
of ODEs, the standard form is
\begin{eqnarray}
\label{stdlinODE}
\dot{T_i} = 
\sum_{j} J_{ij} \,T_{j} + 
\frac{\dot{Q}_i(t)}{C_i}\,,
\quad i= 1,\ldots,N,
\end{eqnarray}
where $J$ (the Jacobian matrix)
is now given by:
{\setlength\arraycolsep{2pt}
\begin{eqnarray}
\label{nJij}
J_{ij} &=&  C_i^{-1} \left(K_{ij} + K^R_{ij}\right),
\quad \mathrm{if}\;i \neq j,\\
\label{nJii}
J_{ii} &=& C_i^{-1} \left[-\sum_{k=1}^N \left(K_{ik} + K^R_{ik}
\right) - K^R_i\right].
\end{eqnarray}
}%

The linear system of Eqs.~(\ref{stdlinODE}) can be solved in the standard way, 
yielding:
\begin{equation}
{T}(t) = \exp\left[t J \right]
\left({T}(0) + \int_0^t \exp\left[-\t J \right] \cdot 
q(\t)\,d\t \right),
\end{equation}
where we have introduced the vector $q(t)$, with components
$q_i(t)=\dot{Q}_i(t)/C_i$.  We can also express the solution in terms of the
driving function $F=q-\langle q \rangle$: 
\begin{eqnarray}
{T}(t) = \exp\left[t J \right] {T}(0) + \int_0^t \exp\left[(t-\t) J \right]
\cdot \left(F(\t) + \langle q \rangle \right) d\t = \\
\exp\left[t J \right] {T}(0) + \int_0^t \exp\left[(t-\t) J \right] 
\cdot F(\t)  d\t + J^{-1}(\exp\left[t J \right] -I) \langle q \rangle
\end{eqnarray}
For large $t$, this solution tends to the periodic stationary solution
\begin{equation}
{T}^\infty(t) = \int_0^\infty \exp\left[\t J \right] 
\cdot F(t-\t)  d\t - J^{-1} \langle q \rangle,
\label{T_lim}
\end{equation}
assuming that $\exp\left[t J \right] \ra 0$ as $t\ra\infty$.  This is a
consequence of the structure of $J$, as in the preceding section.  In the
present case, the eigenvalues of $J$, beyond having negative real parts,
are actually negative real numbers, as we show below.

The total conductance matrix $K + K^R$ is symmetric but this does not imply
that $J$ is symmetric.  Nevertheless, if we define $C=
\mathrm{diag}(C_1,\ldots,C_N)$, the matrix $C^{1/2}\cdot J\cdot C^{-1/2}$ is
symmetric, because its off-diagonal matrix elements are:
$$ 
\left(C^{1/2}\, J\,C^{-1/2}\right)_{ij} =  
\frac{K_{ij} + K^R_{ij}}{\sqrt{C_iC_j}},
\quad i= 1,\ldots,N,\; j= 1,\ldots,N,\;
\mathrm{and}\;i \neq j.\\
$$
Hence, the matrix $C^{1/2} J C^{-1/2}$, similar to $J$, has real eigenvalues.
Furthermore, $C^{1/2} J C^{-1/2}$ is diagonalized by an orthogonal
transformation; that is to say, there is an orthogonal matrix $\O$ such that
$$\O^t\cdot\left(C^{1/2}\, J\,C^{-1/2}\right)\cdot\O =
(C^{-1/2} \O)^{-1} \cdot J\cdot(C^{-1/2} \O)
$$
is diagonal. Therefore, the thermal modes are actually {\em normal}; that is
to say, the modes, which are the eigenvectors of $J$ and hence the columns of
the matrix $P = C^{-1/2} \O$, are related to the eigenvectors of $C^{1/2} J
C^{-1/2}$, which are normal and are given by the columns of $\O$.
Alternatively, one can say that the eigenvectors of $J$ are normal in the
``metric'' defined by $C$; namely,
$$\sum_{i=1}^N C_i \,P_{ia} P_{ib} = \d_{ab}\,,$$
which can be written in matrix form as $P^t C P= I$.
Naturally, the orthogonality of modes greatly simplifies some computations.

Furthermore, the symmetry of the conductance matrix implies that the sum in
Eq.~(\ref{linODE}) can be written as the action of a {\em graph Laplacian}
\cite{Chung} on the temperature vector. Naturally, the graph is formed by the
nodes and the linking conductances. A graph Laplacian is a discretization of
the ordinary Laplacian and is conventionally defined with the sign that makes
it positive semidefinite. The zero eigenvalue corresponds to a constant
function, that is, a constant temperature, in the present case. A vector with
equal components, say, equal to $1/\sqrt{N}$, is the positive (Perron)
eigenvector of the matrix.  With more generality, the Laplacian of a graph can
be defined as a symmetric matrix with off-diagonal entries that are negative
if the nodes are connected and null if they are not \cite{graph_eigenvec}.
This definition does not constrain the diagonal entries and, therefore, does
not imply that a graph Laplacian is positive semidefinite.  It can be made
positive definite (or just semidefinite) by adding to it a multiple of the
identity matrix, which does not alter the eigenvectors. Of course, the
eigenvector corresponding to the smallest eigenvalue does not have to be
constant, but the Perron-Frobenius theorem \cite{Ber-Plem} tells us that it is
positive.  By this general definition of a graph Laplacian, the matrix
$-C^{1/2} J C^{-1/2}$ is a different Laplacian for the same graph, and
Eqs.~(\ref{stdlinODE}) contain the action of this Laplacian on the vector
$C^{1/2} T$.  Notice that this general definition of a graph Laplacian is
connected with the definition of a $Z$-matrix \cite{Ber-Plem} and, actually, a
symmetric $Z$-matrix is a graph Laplacian. If such a matrix is positive
definite, then it is equivalent to a {\em Stieltjes matrix}, namely, a
symmetric nonsingular $M$-matrix \cite{Ber-Plem}.  The general Jacobian
obtained in Sect.~\ref{pert-eq} is also such that $-J$ and also $-C^{1/2} J
C^{-1/2}$ are both nonsingular $M$-matrices, but they need not be symmetric.

To investigate the accuracy of the approximation of the radiation terms by
conduction terms, let us compare the periodic solution given by
Eq.~(\ref{T_lim}) with the first-order perturbative solution found
in Sect.~\ref{pert-eq}, namely, ${T}^\infty(t) = \widetilde{T} + {T}^\infty_{(1)}(t)$.
Of course, the Jacobian matrices in the respective integrals differ, as do
the temperature vectors added to the integrals, namely, $\widetilde{T}$ or
$-J^{-1} \langle q \rangle$.  While $\widetilde{T}$ corresponds to the
authentic steady-state of the nonlinear averaged equations, $-J^{-1} \langle q
\rangle$ corresponds to the steady-state of Eqs.~(\ref{linODE}) after
averaging, which is a state without significance, since we have already
used the set of temperatures $\widetilde{T}$ of the authentic steady-state to
define the radiation conductances $K^R_{ij}$ in Eq.~(\ref{linODE}).
Therefore, the only sensible linear solution is the perturbative solution
${T}^\infty(t) = \widetilde{T} + {T}^\infty_{(1)}(t)$, even if we replace the
Jacobian matrix given by Eq.~(\ref{Jij}) and (\ref{Jii}) with the one given by
Eq.~(\ref{nJij}) and (\ref{nJii}).

In our context, the notion of radiation conductance actually follows from the
symmetry of the matrices $CJ$ or $C^{1/2} J C^{-1/2}$. Therefore, the most
natural definition of radiation conductance probably is $K^R_{ij} = 2 R_{ij}
(\widetilde{T}_{i}^3 + \widetilde{T}_{j}^3)$, that is, the symmetrization of
the term $4 R_{ij} \widetilde{T}_{j}^3$ in Eq.~(\ref{Jij}).  This
symmetrization has been tested by Krishnaprakas \cite{Kris2}, considering the
steady-state problem for models with up to $N=1237$ nodes and working with
various resolution algorithms. He found that the effect of symmetrization is
not appreciable.  To estimate the effect of the antisymmetric part of the
matrix $4 R_{ij}\widetilde{T}_{j}^3$, namely, $2 R_{ij} (\widetilde{T}_{i}^3 -
\widetilde{T}_{j}^3)$, on the eigenvalue problem for the Jacobian, we proceed
as follows.  We formulate this eigenvalue problem in terms of the matrix
$C^{1/2} J C^{-1/2}$, so that it is an eigenvalue problem for a symmetric
matrix perturbed by a small antisymmetric part.  This problem is well
conditioned, because the eigenvectors of the symmetric matrix (the columns of
the matrix $\O$) are orthogonal.  In particular, the perturbed eigenvalues are
still real. Furthermore, the first-order perturbation formula for the
eigenvalue $\l_a$ associated with an eigenvector $e_a$ \cite{pert-methods}
yields:
$$
\d \l_a = 
\sum_{i,j=1}^N \d A_{ij}\, e_{ai} \, e_{aj}
= 0, 
$$ 
vanishing because the perturbation matrix $\d A$ is antisymmetric. So the
nonvanishing perturbative corrections begin at the second order in the
perturbation matrix, and, in this sense, they are especially small.

\section{Fourier analysis of the periodic solution}
\label{Fourier-anal}

Given that ${T}_{(1)}^\infty(t)$ is a periodic function, it can be expanded in
a Fourier series.  To derive this series, let us first introduce the Fourier
series of $F(t)$,
$$F(t) = \sum_{m = -\infty}^{\infty}\hat{F}(m)\, e^{2\pi i m t/\T}.$$
Inserting this series in the integral of Eq.~(\ref{T_1_lim}) and integrating
term by term, we obtain the Fourier series for ${T}_{(1)}^\infty(t)$.
Alternatively, we can substitute the Fourier series for both
${T}_{(1)}^\infty(t)$ and $F(t)$ into Eqs.~(\ref{linODE-pert}), where $J$ is
taken to be constant; then we can solve for the Fourier coefficients of
${T}_{(1)}^\infty(t)$. The result is
\begin{equation}
{T}_{(1)}^\infty(t) = \sum_{m = -\infty}^{\infty} 
e^{2\pi i m t/\T} \left(2\pi i m I/\T - J\right)^{-1} \cdot \hat{F}(m)\,, 
\label{Fourier-sol}
\end{equation}
The Fourier coefficients $\hat{F}(m)$ are obtained by integration:
\begin{equation}
\hat{F}(m) = \frac{1}{\T}\int_{0}^{\T} F(t)\, e^{-2\pi i m t/\T}\,dt\,.
\label{Fourier-coef}
\end{equation}
Given that $F(t)$ is a real function, 
\begin{equation}
\hat{F}(-m)=\hat{F}^*(m).
\label{F*}
\end{equation}
Furthermore, $\langle F(t) \rangle = 0$ implies 
\begin{equation}
\hat{F}(0) = 0.
\label{F0}
\end{equation}
So $F(t)$ is defined by the sequence of Fourier coefficients for positive
$m$. This sequence must fulfill the requirement that $\lim_{m \ra
  \infty}\hat{F}(m) = 0,$ so a limited number of the initial coefficients may
suffice.

Actually, for numerical work, Eq.~(\ref{Fourier-coef}) can be conveniently
replaced by the discrete Fourier transform
\begin{equation}
\hat{F}(m) = \frac{1}{n}\sum_{k=0}^{n-1} 
F(k\T/n) \,e^{-2\pi i m k/n},
\label{d_Fourier-coef}
\end{equation}
which only requires sampling of the values for $F(t)$, but also only defines a
finite number of independent Fourier coefficients, because
$\hat{F}(m+n)=\hat{F}(m)$.  Notice that we usually have available just a
sampling of the heat inputs at regular time intervals, rather than the
analytical form of $\dot{Q}_i(t)$.  To calculate the exact number of
independent Fourier coefficients provided by Eq.~(\ref{d_Fourier-coef}), we
must take into account Eqs.~(\ref{F*}) and (\ref{F0}).  If $n$ is an odd
number, the independent Fourier coefficients $\hat{F}(m)$ are the ones with $m
= 1, \ldots, (n-1)/2$; that is to say, there are $n-1$ independent real
numbers.  If $n$ is even, the independent Fourier coefficients
are the ones with $m = 1, \ldots, n/2$, and 
$$\hat{F}(n/2) = \frac{1}{n}\sum_{k=0}^{n-1} (-)^k F(k\T/n)$$ is real, so
there are $n-1$ independent real numbers as well.  For definiteness, let $n$
be odd. Then, we can express $F(t)$ as
$$F(t) = 
2 \,{\rm Re} \left[\sum_{m = 1}^{(n-1)/2}\hat{F}(m)\, e^{2\pi i m t/\T}\right].
$$
Of course, the values of $F(t)$ at $t=k\T/n, \; k=0,\ldots,n-1,$ are the
sampled values employed in Eq.~(\ref{d_Fourier-coef}), but the expression is
valid for any $t \in [0,\T]$ and constitutes an interpolation of the sampled
values.  Naturally, the higher the sampling frequency $n$, the more
independent Fourier coefficients we have and the more accurate the
representation of $F(t)$ is.

As is well known, the Fourier series of a function $F(t)$ that is
piecewise smooth converges to the function, except at its points of
discontinuity, where it converges to the arithmetic mean of the two one-sided
limits \cite{Fourier}.  However, the convergence is not {\em uniform}, so that
partial sums oscillate about the true value of the function near each point
of discontinuity and ``overshoot'' the two one-sided limits in opposite
directions. This overshooting is known as the Gibbs phenomenon, and, in our
case, produces typical errors near the discontinuities of the driving function
$F$. These discontinuities are due to the sudden obstructions of the radiation
on parts of the aircraft that occur at certain orbital positions, for example,
when the Sun is eclipsed.%
\footnote{Strictly speaking, the function $\dot{Q}_i(t)$ is always 
continuous but it undergoes sharp variations at some times. These sharp
variations can be considered as discontinuities, especially, if the function
is sampled.}
Section \ref{DNS} shows that the Gibbs phenomenon at eclipse points
can be responsible for the largest part of the error 
of the linear method when the discrete Fourier transform is used.

The approximation of ${T}_{(1)}^\infty(t)$ provided by the $n$ samples of
$F(t)$ is, of course,
\begin{equation}
{T}_{(1)}^\infty(t) = 2 \,{\rm Re} \left[\sum_{m = 1}^{(n-1)/2}
e^{2\pi i m t/\T} \left(2\pi i m I/\T - J\right)^{-1} \cdot \hat{F}(m)\right],
\label{d_Fourier-sol}
\end{equation}
and is valid for any $t \in [0,\T]$. However, if we are only interested in
$T_{(1)}^\infty(t)$ at $t=k\T/n$, $k=0,\ldots,n-1,$ 
we can compute these values with the inverse discrete Fourier transform
\begin{equation}
{T}_{(1)}^\infty(k\T/n) = 
\sum_{m = 0}^{n-1}
e^{2\pi i m k/n} \left(2\pi i 
\left[\mathrm{mod}\left(m+\frac{n-1}{2}\,,n\right)-\frac{n-1}{2}\right] I/\T -
J\right)^{-1} \cdot \hat{F}(m)\,, 
\label{i_d_Fourier-tr}
\end{equation}
where, for $m = (n+1)/2, \ldots,n-1,$ $\hat{F}(m) = \hat{F}(m-n) =
\hat{F}^*(n-m)$, and where $\mathrm{mod}\left(\cdot,n\right)$ gives the
remainder of the integer division by $n$. This inverse discrete Fourier
transform can be more convenient for a fast numerical computation.  Regarding
computational convenience, the discrete Fourier transform, be it direct or
inverse, is best performed with a fast Fourier transform (FFT) algorithm. The
classic FFT algorithm requires $n$ to be a power of two
\cite{Num_rec,Gol-Van}; in particular, it has to be even.

The function ${T}_{(1)}^\infty(t)$, computed by Fourier analysis from $n$
samples of $\dot{Q}_i(t)$, is to be compared with the one computed by a
numerical approximation of the integral formula, Eq.~(\ref{T_1_lim}), in terms
of the same samples.  Naturally, we can use instead of the integral over $\t
\in [0,\infty]$ the integral over $\t \in [0,\T]$ below
Eq.~(\ref{T_1_lim}). This integral can be computed from the $n$ samples of
$F(t)$ by an interpolation formula, say the trapezoidal rule.  It is not easy
to decide whether this procedure is more efficient than Fourier transforms.
Considering that the substitution of the continuous Fourier transform,
Eq.~(\ref{Fourier-coef}), by the discrete transform,
Eq.~(\ref{d_Fourier-coef}), is equivalent to computing the former with the
trapezoidal rule, the integral formula may seem more direct. In particular,
this formula allows us to select the values of $t$ for which we compute
${T}_{(1)}^\infty(t)$ independent of the sampling frequency, so we can choose
just a few distinguished orbital positions and avoid the computation of all
the $n-1$ integrals (one is removed by the condition $\langle
{T}_{(1)}^\infty(t) \rangle = 0$).  Note that the computation of all of the
independent $\hat{F}(m)$ with Eq.~(\ref{d_Fourier-coef}) is equivalent to the
computation of precisely $n-1$ integrals.  However, the efficiency of the FFT
reduces the natural operation count of this computation, of order $n^2$, to
order $n \log n$; so its use can be advantageous, nevertheless.

It goes without saying that the second-order perturbative contribution
${T}_{(2)}^\infty(t)$ to the stationary solution is given by the right-hand
side of Eq.~(\ref{d_Fourier-sol}) with the Fourier coefficients $\hat{F}(m)$
replaced by the Fourier coefficients of the function $G(t)$ defined in
Sect.~\ref{second-order}.

\section{Ten-node model of a Moon-orbiting satellite}
\label{ten-node}

\begin{figure}
\centering{\includegraphics[width=10cm]{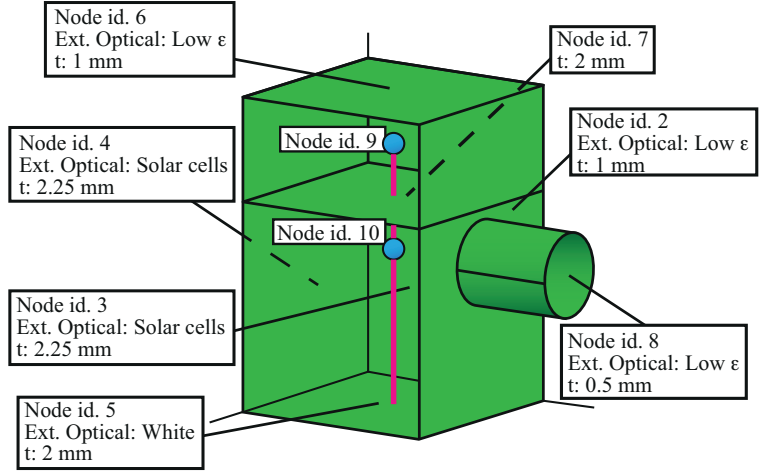}}
\caption{Satellite's structure and node description. The front face, removed to
see the interior, corresponds to node 1 and is equivalent to nodes 3 or 4.}
\label{sat}
\end{figure}

To test the previously explained methods, we construct a small thermal model
of a simple spacecraft, namely, a ten-node model of a Moon-orbiting
satellite. Our satellite ten-node model supports a basic thermal structure and
is simple enough for allowing one to explicitly display the main mathematical
entities, e.g., the matrices $K$, $R$ and $J$.  The satellite consists of a
rectangular parallelepiped (a cuboid) of square base plus a small cylinder on
one of its sides that simulates an observation instrument, as represented in
Fig.~\ref{sat}.  In addition, at a height of two thirds of the total height,
there is an inner tray with the electronic equipment.  The dimensions of the
cuboid are $0.2\,\mathrm{m} \times 0.2\,\mathrm{m} \times 0.3\,\mathrm{m}$,
and the cylinder has a length of 0.1 m and a radius of 0.04 m.  The
satellite's frame is made of aluminum alloy, using plates 1 mm thick, except
the bottom plate, which is 2 mm thick.  This plate plays the role of a
radiator and its outer surface is painted white to have high solar
reflectance.  The cylinder is made of the same aluminum alloy, as well as the
tray; they are 0.5 mm and 2 mm thick, respectively.  The sides of the
satellite, except the one with the instrument, are covered with solar cells,
which increase the sides' thickness to 2.25 mm.

The thermal model of the satellite assigns one node to each face of the
cuboid, one more to the cylinder and another to the tray, that is, eight nodes
altogether. Furthermore, to conveniently split the total heat capacitance of
the electronic equipment, it is convenient to add two extra nodes with (large)
heat capacitance but with no surface that could exchange heat by radiation.
Nodes of this type are called ``non-geometrical nodes''.  In the present case,
they represent two boxes with equipment placed above and below the tray,
respectively.  We order the ten nodes as shown in Fig.~\ref{sat}.  The lower
box (node 10) is connected to the radiator by a thermal strap.  Given the
satellite's structure and assuming appropriate values of the specific heat
capacities, it is possible to compute the capacitances $C_i, \; i=1\ldots,10,$
with the result given in Table~\ref{tab1}.  Using the value of the
aluminum-alloy heat conductivity and assuming perfect contact between plates,
we compute the conduction coupling constants $K_{ij}$ between nodes $i,j=
1,\ldots,8$. The remaining conduction coupling constants are given reasonable
values, shown in Eq.~(\ref{Kij}).  The computation of the radiation coupling
constants $R_{ij},\; i,j= 1,\ldots,8,$ and $R_{i},\; i= 1,\ldots,8,$ and
indeed the computation of the external radiation heat inputs requires a
detailed {\em radiative model} of the satellite, consisting of the geometrical
{\em view factors} and the detailed thermo-optical properties of all
surfaces. This radiative model allows us to compute the respective {\em
  absorption factors} \cite{Kreith}.

\begin{table}
\begin{center}
\begin{tabular}{r|r|r|r}
Node & $C_i$ (J/K) &
$\langle{\dot{Q_i}}\rangle$ (W) & $\widetilde{T}_i$ ($^{\circ}\mathrm{C}$)\\
\hline
 1\phantom{a} & 331.7\phantom{a} & 15.18\phantom{a} & 2.6\phantom{a} \\
 2\phantom{a} & 147.4\phantom{a} & 2.30\phantom{a} & 3.6\phantom{a} \\
 3\phantom{a} & 331.7\phantom{a} & 15.17\phantom{a} & 2.6\phantom{a} \\
 4\phantom{a} & 331.7\phantom{a} & 14.80\phantom{a} & 2.3\phantom{a} \\
 5\phantom{a} & 196.6\phantom{a} & 3.91\phantom{a} & 0.2\phantom{a} \\
 6\phantom{a} & 98.3\phantom{a} & 0.63\phantom{a} & 2.2\phantom{a} \\
 7\phantom{a} & 196.6\phantom{a} & 0\phantom{aa} & 6.3\phantom{a} \\
 8\phantom{a} & 31.9\phantom{a} & 1.70\phantom{a} & 4.7\phantom{a} \\
 9\phantom{a} & 800.0\phantom{a} & 4.35\phantom{a} & 15.9\phantom{a} \\
 10\phantom{a} & 1400.0\phantom{a} & 6.15\phantom{a} & 11.1\phantom{a}
\end{tabular}
\end{center}
\caption{Node capacities and mean heat inputs with their
associated steady-state temperatures.}
\label{tab1}
\end{table}

\begin{figure}
\centering{\includegraphics[width=6cm]{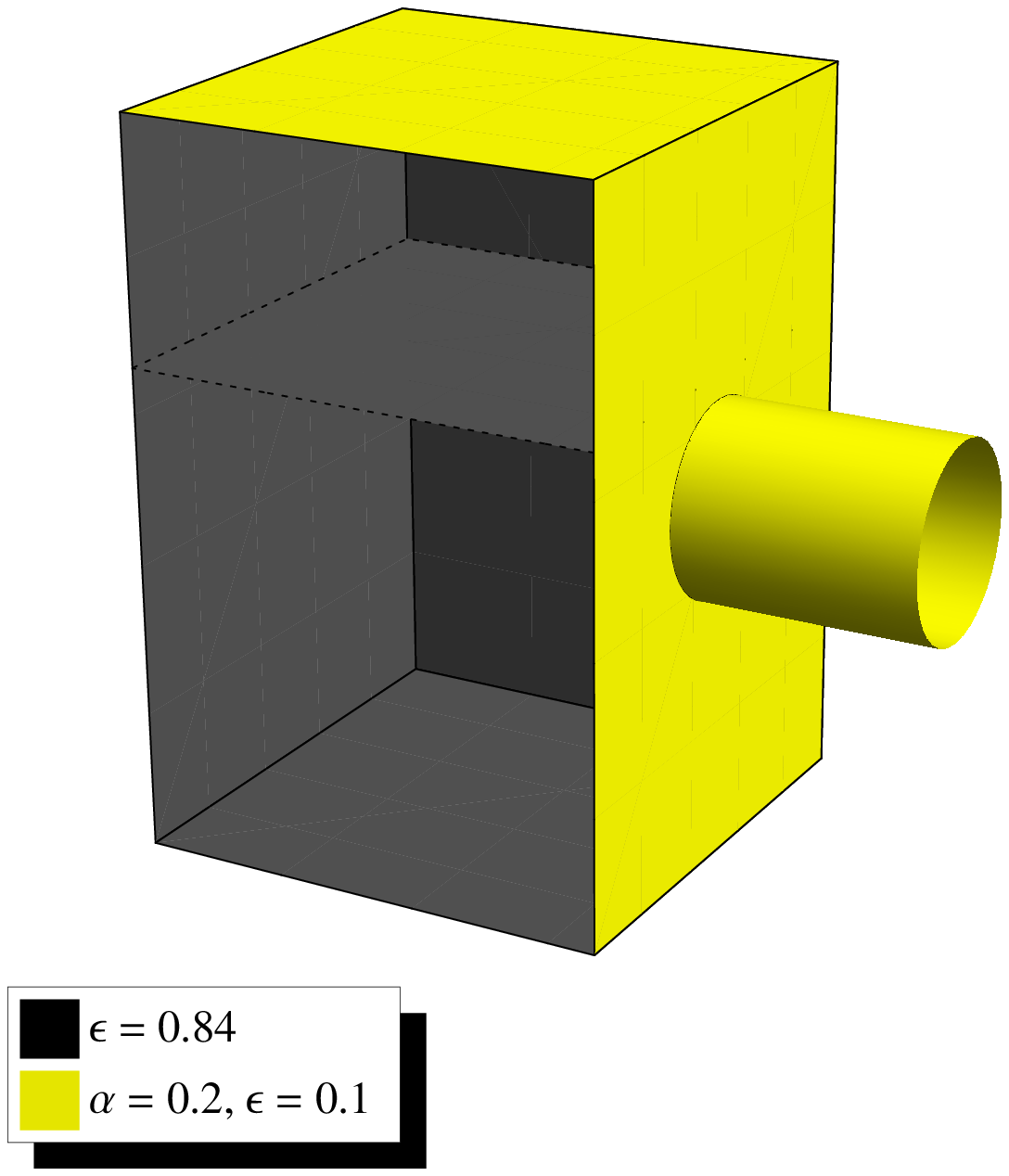}}
\centering{\includegraphics[width=6cm]{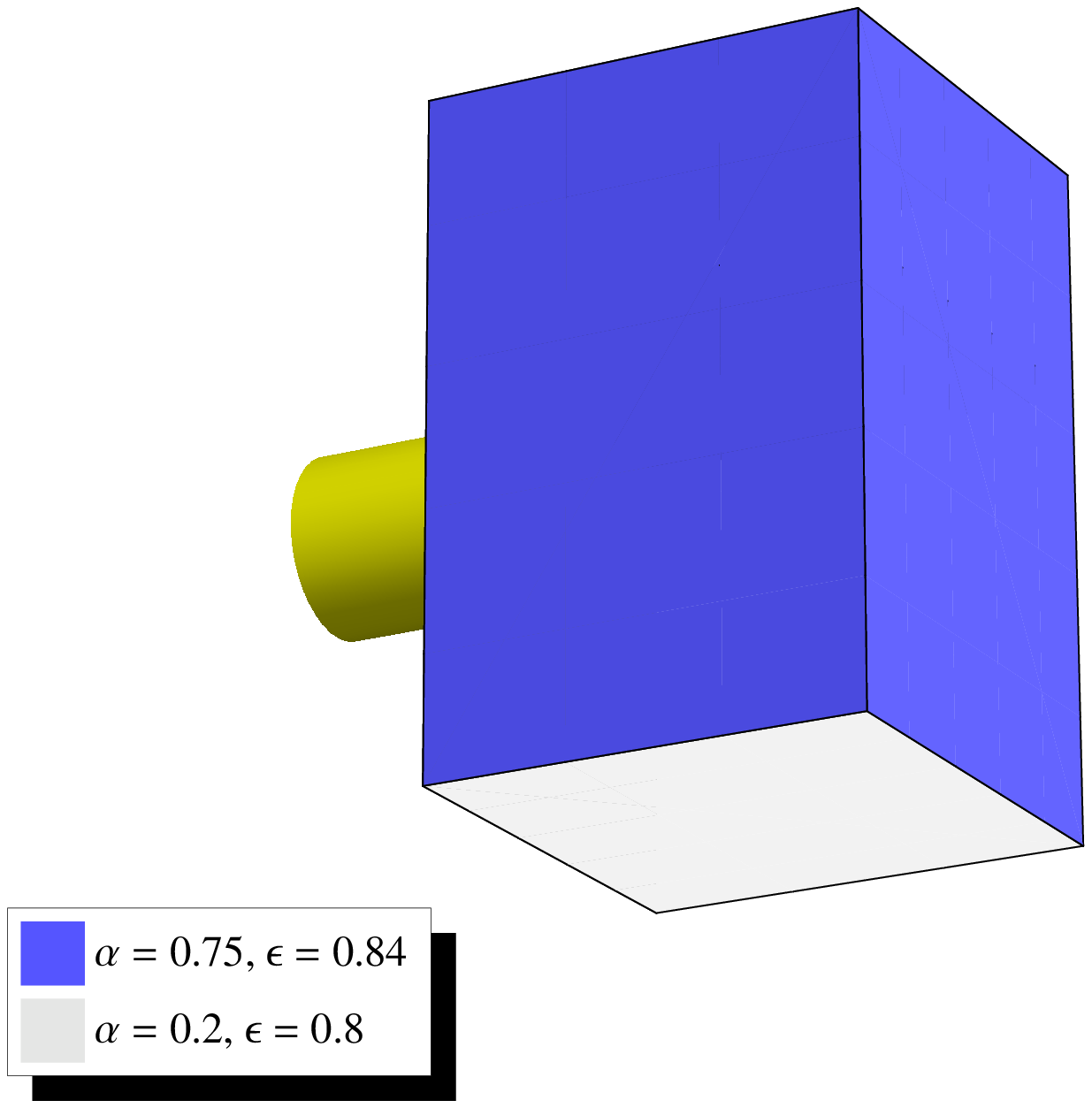}}
\caption{Thermo-optical properties of the satellite's surfaces (details are given in
the text).}
\label{opt}
\end{figure}

The thermo-optical properties of the surfaces are assumed to be as realistic
as possible, given the simplicity of the thermal model.  All radiation
reflection is assumed to be diffuse, as is common for many types of
surfaces. The inner surfaces are painted black and have high emissivity, $\ee
= 0.84$, to favor the uniformization of the interior temperature. The outer
surfaces are of three types. The three sides covered with solar cells also
have high emissivity, $\ee = 0.84$, to favor the cooling of the solar
cells. On the other hand, they have high solar absorptivity, $\a_\mathrm{s} =
0.75$. Of this 0.75, 0.18 is processed into electricity and the remaining 0.57
dissipates as heat in the solar cells.  The top surface, the surface with the
cylinder, and the cylinder itself (its two sides) have low emissivity, $\ee =
0.1$, and low solar absorptivity, $\a_\mathrm{s} = 0.2$, which are chosen to
simulate the effect of a {\em multilayer insulator}.  In contrast, the bottom
surface simulates a radiator, with $\ee = 0.8$ and $\a_\mathrm{s} = 0.2$ (like
an {\em optical solar reflector}).  All of these thermo-optical properties are
summarized in Fig.~\ref{opt}.  For the computation of the corresponding
absorption factors, we employ the ray-tracing Monte-Carlo simulation method
provided by ESARAD$^\mathrm{TM}$ (ESATAN$^\mathrm{TM}$'s radiation module)
\cite{ESATAN}.

Taking into account the above information, one obtains the following
conduction (in W/K) and radiation (in W/K$^4$) matrices:
\begin{eqnarray}
(K_{ij}) &=&
\frac{1}{10}
\left(
\begin{array}{cccccccccc}
 0 & 3.47 & 0 & 5.64 & 2.86 & 2.00 & 4.50 & 0 & 0 & 0 \\
 3.47 & 0 & 3.47 & 0 & 1.67 & 1.33 & 3.50 & 3.00 & 0 & 0 \\
 0 & 3.47 & 0 & 5.64 & 2.86 & 2.00 & 4.50 & 0 & 0 & 0 \\
 5.64 & 0 & 5.64 & 0 & 2.86 & 2.00 & 4.50 & 0 & 0 & 0 \\
 2.86 & 1.67 & 2.86 & 2.86 & 0 & 0 & 0 & 0 & 0 & 3.00 \\
 2.00 & 1.33 & 2.00 & 2.00 & 0 & 0 & 0 & 0 & 0 & 0 \\
 4.50 & 3.50 & 4.50 & 4.50 & 0 & 0 & 0 & 0 & 4.50 & 6.00 \\
 0 & 3.00 & 0 & 0 & 0 & 0 & 0 & 0 & 0 & 0 \\
 0 & 0 & 0 & 0 & 0 & 0 & 4.50 & 0 & 0 & 0 \\
 0 & 0 & 0 & 0 & 3.00 & 0 & 6.00 & 0 & 0 & 0
\end{array}
\right),
\label{Kij}
\\
(R_{ij}) &=& 
10^{-10}
\left(
\begin{array}{cccccccccc}
 0 & 5.06 & 4.63 & 5.05 & 3.68 & 2.71 & 6.39 & 0 & 0 & 0 \\
 5.06 & 0 & 5.05 & 4.63 & 3.68 & 2.70 & 6.39 & 0.13 & 0 & 0 \\
 4.63 & 5.05 & 0 & 5.06 & 3.69 & 2.71 & 6.39 & 0 & 0 & 0 \\
 5.05 & 4.63 & 5.06 & 0 & 3.69 & 2.70 & 6.38 & 0 & 0 & 0 \\
 3.68 & 3.68 & 3.69 & 3.69 & 0 & 0 & 3.57 & 0 & 0 & 0 \\
 2.71 & 2.70 & 2.71 & 2.70 & 0 & 0 & 7.19 & 0 & 0 & 0 \\
 6.39 & 6.39 & 6.39 & 6.38 & 3.57 & 7.19 & 0 & 0 & 0 & 0 \\
 0 & 0.13 & 0 & 0 & 0 & 0 & 0 & 0 & 0 & 0 \\
 0 & 0 & 0 & 0 & 0 & 0 & 0 & 0 & 0 & 0 \\
 0 & 0 & 0 & 0 & 0 & 0 & 0 & 0 & 0 & 0
\end{array}
\right),
\label{Rij}
\\
(R_{i}) &=& 10^{-9}\,(2.86,0.32,2.86,2.86,1.81,0.23,0,0.23,0,0).
\label{Ri}
\end{eqnarray}

The satellite's thermal characteristics are defined by the data set $\{C_i,
K_{ij}, R_{ij}, R_{i}\},$ but the radiation heat exchange depends on the nodal
temperatures, which in turn depend on the heat input.  As explained in
Sect.~\ref{pert-eq}, the appropriate set of nodal temperatures corresponds to
the steady-state for averaged heat inputs, given by the algebraic equation
that results from making $\dot{T}_{(0)i} = 0$ in Eq.~(\ref{bODE-pert}).  Since
we need the external heat inputs and, therefore, the orbit, we proceed to
define the orbit characteristics.

\begin{figure}
\centering{\includegraphics[width=8cm]{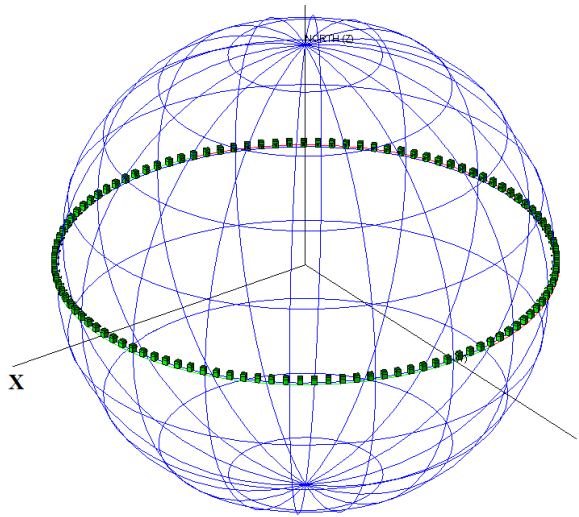}}
\caption{The 111 positions of the satellite in its orbit. 
The sunlight comes along the $x$ axis.}
\label{orbit}
\end{figure}

We choose a circular equatorial orbit $26\hspace{1pt}926$ m above the Moon's
surface, such that $\T = 6660$ s. The radiation heat input to the satellite
consists, on the one hand, of the solar irradiation and the Moon's albedo,
and, on the other hand, of the Moon's constant IR radiation.  We take 0.12 for
the mean Moon's albedo and 270 K for the black-body equivalent temperature of
the Moon.  There is also heat produced by the dissipation of electrical power in
the equipment (nodes 9 and 10).  For the sake of simplicity, the dissipation
rate is assumed to be constant, equal to the mean electrical power generated in
an orbit.  In a part of the orbit, the Moon eclipses the Sun, so the satellite
receives no direct sunlight or albedo, although there is always IR radiation
from the Moon. The satellite is stabilized such that the cylinder (the
``observation instrument'') always points to the Moon and the longer edges are
perpendicular to the orbit. The radiation heat input can be computed by
taking into account the given orbital characteristics and the satellite's
thermo-optical characteristics, in particular, the absorption factors. It
has been computed with ESARAD$^\mathrm{TM}$, taking 111 positions on the
orbit, that is, at intervals of one minute.

\begin{figure}
\begin{minipage}{8.4cm}
\centering{\includegraphics[width=8cm]{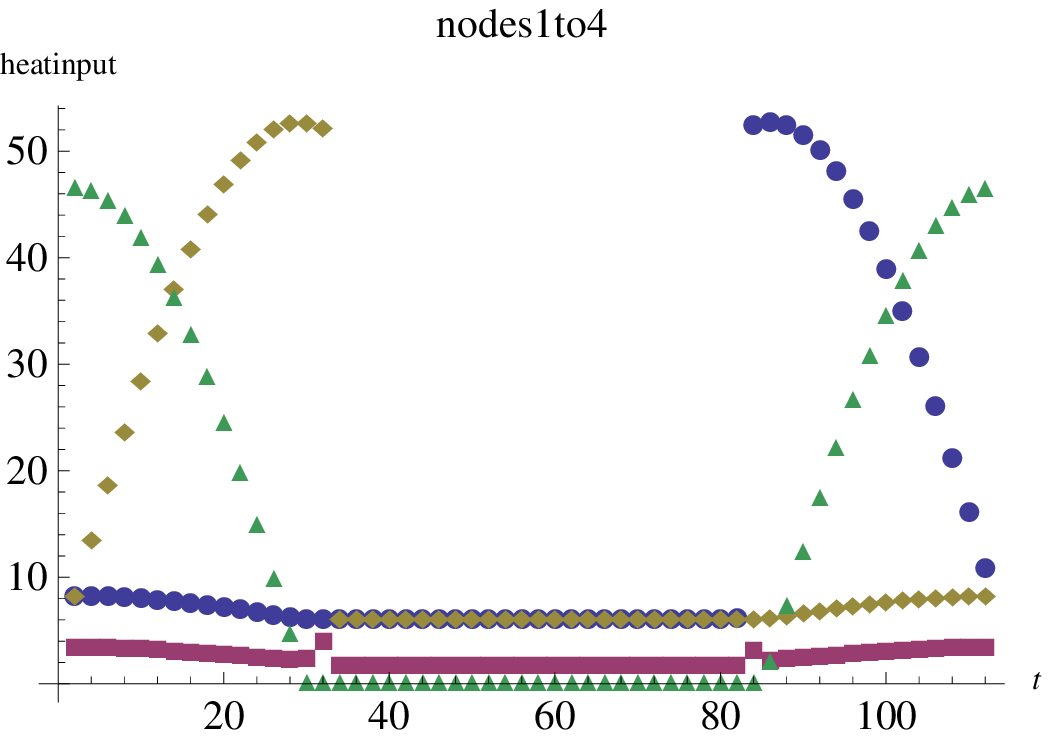}\\(a) nodes 1--4}
\end{minipage}
\begin{minipage}{8.4cm}
\centering{\includegraphics[width=8cm]{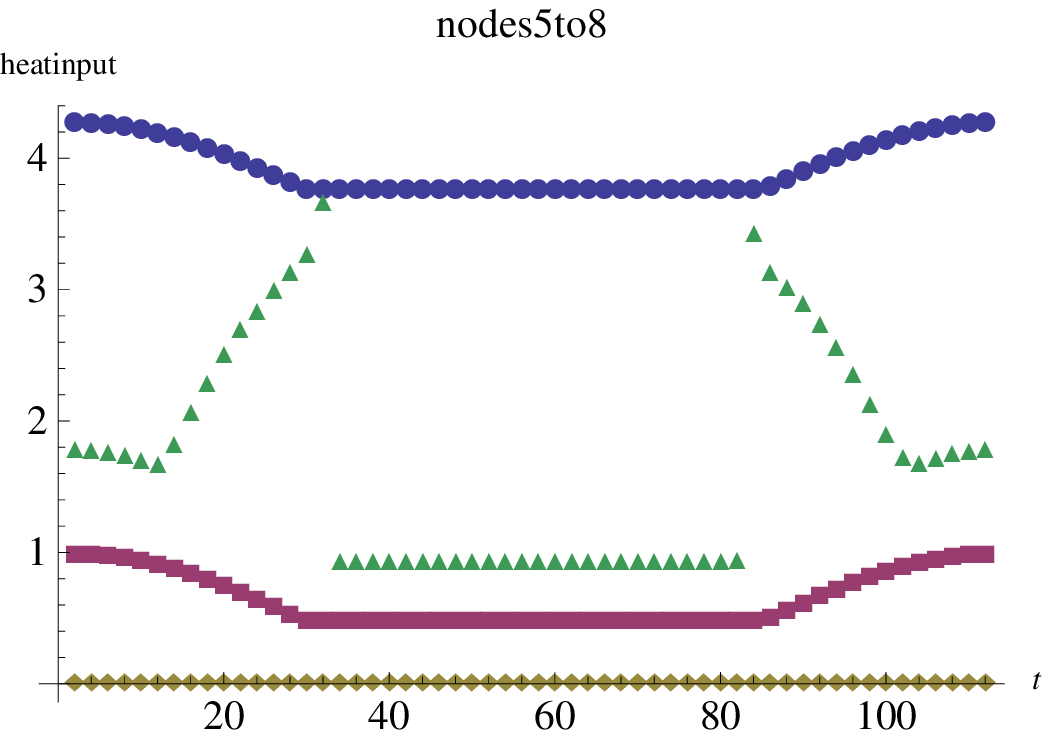}\\(b) nodes 5--8}
\end{minipage}
\\[2mm]
\caption{Variation of the external heat input with time $t$ (in minutes) along the orbit.
For both plots, the node numbers are denoted by shape, in the order: dots
(1,5), squares (2,6), diamonds (3,7), and triangles (4,8).}
\label{heat-in}
\end{figure}

In Fig.~\ref{orbit}, all 111 positions are plotted.  The initial position of
the satellite is at the subsolar point and it moves towards the east.  The
total external radiation heat input to the first eight nodes (the ones that
receive radiation) is plotted in Fig.~\ref{heat-in} (only at every other
position, for clarity).  Note the symmetry between nodes 1 and 3, which denote
the lateral faces, covered with solar cells.  Node 4 corresponds to the back
side, also covered with solar cells. So the external radiation load on it has
a similar time variation, but it is displaced.  The solar radiation absorbed
by all solar cells results in an orbital mean power rate of 10.5 W, dissipated
in the equipment and split between nodes 9 and 10, which receive 4.35 and 6.15
W, respectively.  The external radiation absorbed by the side with the
cylinder (node 2) is considerably smaller than the radiation absorbed by the
sides with solar cells, due to the low value of $\a_\mathrm{s}$ (and of $\ee$,
as well) for the corresponding surface.  The bottom and top outer surfaces,
which belong to nodes 5 and 6, respectively, have view factors for the
external radiation that are much less favorable than those of the side
surfaces. Nevertheless, the amount of lunar IR radiation absorbed by the
bottom surface, due to its high $\ee$, is such that the orbital mean of the
external heat input to node 5 is, in fact, larger than the one for node 2 (see
Table~\ref{tab1}).  Naturally, node 7, with no outer surfaces, does not absorb
any external radiation.

\psfrag{heatinput}{{$\dot{Q}_i$ (W)}}
\psfrag{nodes1to4}{{}}
\psfrag{nodes5to8}{{}}
\psfrag{t}{{$t$}}

To determine the hot and cold cases of the orbit, we compute the total heat
load on the satellite for each position in the orbit, finding a maximum of
90.59~W at position 14 and a minimum of 18.65~W at any position in the
eclipse, during which all the heat loads stay constant.  The solution of the
corresponding steady-state problems at position 14 and at a position in the
eclipse yields the two sets of nodal temperatures (for the given node order):
\begin{description}
\item[Hot]$\{29.0,35.5,41.2,38.5,30.2,35.1,39.1,35.0,48.7,42.9\}\;
^{\circ}\mathrm{C}$. 
\item[Cold]$\{-46.8,-45.0,-46.8,-49.1,-45.7,-47.1,-42.8,-44.1,-33.2,-37.0\}
\;^{\circ}\mathrm{C}$.
\end{description}
The results in Sect.~\ref{J&p-sol} show that the periodic thermal state does
not reach these extreme temperatures, which could endanger the performance of
the satellite.

\subsection{Jacobian matrix and periodic solution}
\label{J&p-sol}

We compute the averages $\langle{\dot{Q_i}}\rangle$ and substitute for them
and the data set $\{K_{ij}, R_{ij},R_{i}\}$ in Eq.~(\ref{bODE-pert}) to find
the steady state temperatures.  The results are given in
Table~\ref{tab1}. Then, according to Eqs.~(\ref{Jij}) and (\ref{Jii}), the
Jacobian matrix is (in s$^{-1}$)
$$
{\setlength\arraycolsep{3pt}
J =
10^{-3}
\left(
\begin{array}{cccccccccc}
 -6.99 & 1.18 & 0.12 & 1.83 & 0.95 & 0.67 & 1.52 & 0 & 0 & 0 \\
 2.64 & -12.93 & 2.64 & 0.26 & 1.33 & 1.06 & 2.75 & 2.04 & 0 & 0 \\
 0.12 & 1.17 & -6.99 & 1.83 & 0.95 & 0.67 & 1.52 & 0 & 0 & 0 \\
 1.83 & 0.12 & 1.83 & -7.64 & 0.95 & 0.67 & 1.52 & 0 & 0 & 0 \\
 1.61 & 1.01 & 1.61 & 1.61 & -8.26 & 0 & 0.16 & 0 & 0 & 1.53 \\
 2.27 & 1.59 & 2.27 & 2.27 & 0 & -9.20 & 0.64 & 0 & 0 & 0 \\
 2.56 & 2.06 & 2.56 & 2.56 & 0.15 & 0.31 & -15.60 & 0 & 2.29 & 3.05 \\
 0 & 9.43 & 0 & 0 & 0 & 0 & 0 & -10.05 & 0 & 0 \\
 0 & 0 & 0 & 0 & 0 & 0 & 0.56 & 0 & -0.56 & 0 \\
 0 & 0 & 0 & 0 & 0.21 & 0 & 0.43 & 0 & 0 & -0.64
\end{array}
\right).
}
$$
By inspection, one can check that it has nonnegative off-diagonal and negative
diagonal elements, that is to say,
$-J$ is a $Z$-matrix. 
It is also {\em diagonally dominant}, namely, $|J_{ii}| \geq \sum_{j \neq i}
|J_{ij}|$.
The eigenvalues of $J$ are
$$
-10^{-4}
\{182.20,154.30,103.40,98.03,86.12,71.09,71.04,14.90,5.70,1.72\}.
$$
Their inverses (in absolute value) give us the typical relaxation times of the
corresponding thermal modes.  Thus, we deduce that relaxation time of the
fastest mode is about 55 s, whereas the relaxation time of the slowest one is
$5\hspace{1pt}813$ s.  The latter time is similar to $\T =
6\hspace{1pt}660$~s.

The eigenvalues are real numbers and, furthermore, $J$ is diagonalizable,
because the eigenvalues are different.  Both properties also follow from
$C^{1/2} J C^{-1/2}$ being almost symmetric: its antisymmetric part, $\d A =
(C^{1/2} J C^{-1/2} - C^{-1/2} J^t C^{1/2})/2$, is relatively small, namely,
$\|\d A\|/\|C^{1/2} J C^{-1/2}\| < 10^{-3}$, where the matrix norm is the
Frobenius norm (other standard matrix norms yield similar values).  Therefore,
the notion of ``radiation conductance'' (Sect.~\ref{heur}) is appropriate in
this case, as concerns its use in the linear equations.  The thermal modes are
almost normal, namely, the eigenvector matrix $P$ is such that $P^t C P= I$
with an error $< 0.002.$ The most interesting eigenvector of $J$ is, of
course, the positive (Perron) eigenvector, which corresponds to the slowest
mode.  The normalized positive eigenvector is
$$
 (0.259,0.276,0.259,0.257,0.275,0.267,0.327,0.264,0.471,0.423)~\mathrm{K}.
$$
Note that the temperature increments are of a similar magnitude, except the
ones of node 7 and, especially, nodes 9 and 10, which are associated,
respectively, for the tray and the boxes of electronic equipment.  The next
mode, corresponding to the eigenvalue $-5.70\cdot 10^{-4}$, has one negative
component (the ninth), and the remaining modes have more than one.

\begin{figure}
\begin{minipage}{8.4cm}
\centering{\includegraphics[width=8cm]{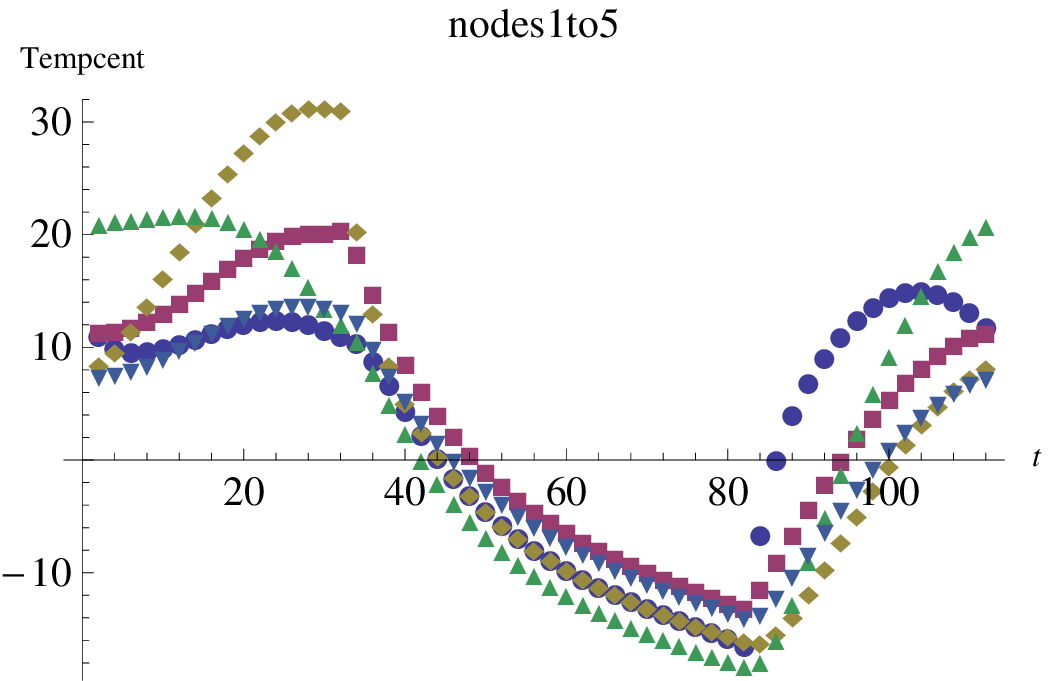}\\(a) nodes 1--5}
\end{minipage}
\begin{minipage}{8.4cm}
\centering{\includegraphics[width=8cm]{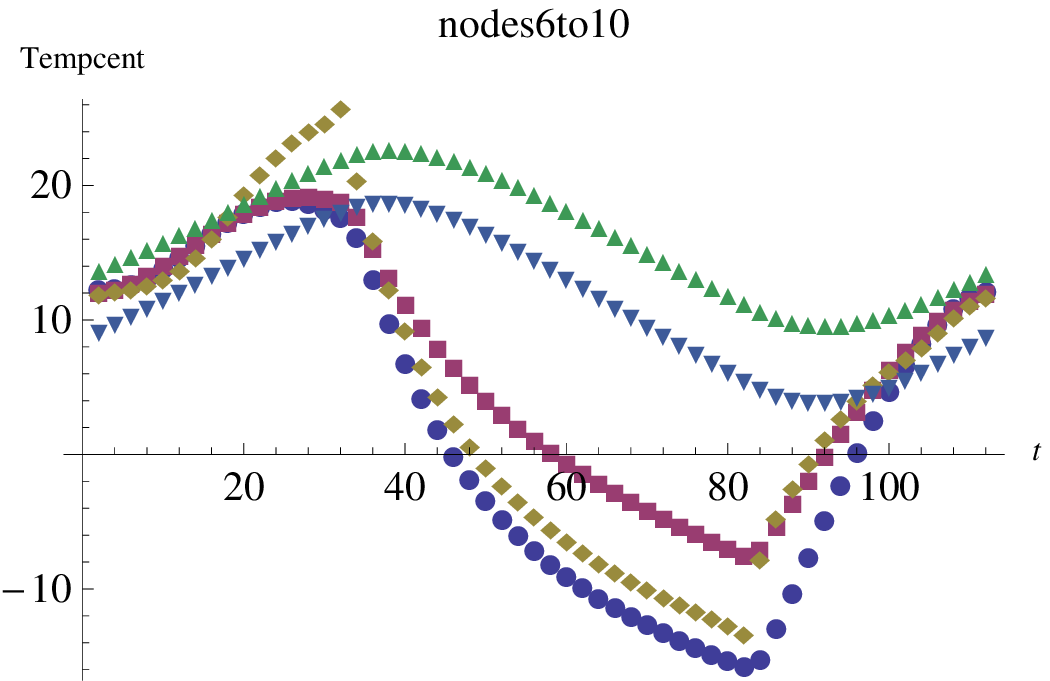}\\(b) nodes 6--10}
\end{minipage}
\\[2mm]
\caption{Variation of the ten nodal temperatures with time $t$ (in minutes).
For both plots, the node order is: 
dots (1,6), squares (2,7), diamonds (3,8),
upward triangles (4,9), and downward triangles (5,10).}
\label{Tevol}
\end{figure}

To calculate ${T}_{(1)}^\infty(t)$, we choose the Fourier series of
Eq.~(\ref{d_Fourier-sol}) or, rather, the inverse discrete Fourier transform
of Eq.~(\ref{i_d_Fourier-tr}), which can be computed with a FFT algorithm.
The Fourier coefficients $\hat{F}(m)$ can also be computed with the FFT,
according to Eq.~(\ref{d_Fourier-coef}).  Once the vector ${T}_{(1)}^\infty$
at the 111 positions is available, the set of nodal temperatures corresponding
to the first-order perturbative solution is ${T}_i^\infty(t) = \widetilde{T}_i
+ {T}^\infty_{(1)i}(t),\,i=1,\ldots,10,$ plotted in Fig.~\ref{Tevol}.  A
measure of the accuracy of this perturbative calculation is given by the
second-order calculation in the next section.  The truncation of the Fourier
series imposed by the sampling of $F$ also is a source of error, unrelated to
perturbation theory.  The piecewise smoothness of the function
${T}_{(1)}^\infty(t)$ suggests that the error is small (but see
Sect.~\ref{DNS}).

\psfrag{Tempcent}{{${T}_i^\infty$ ($^{\circ}\mathrm{C}$)}}
\psfrag{nodes1to5}{{}}
\psfrag{nodes6to10}{{}}

It is also interesting to see if the first-order perturbative calculation is
affected by neglecting the fastest modes: according to the analysis at the end
of Sect.~\ref{pert-eq}, these modes are expected to contribute in proportion
to their relaxation times. The fastest mode relaxes in about 55 s, a short but
non-negligible time. As a consequence, its contribution to ${T}_{(1)}^\infty$,
which we find to have a maximum magnitude of 0.8 K, is small but
non-negligible. But we can deduce that still faster modes, which would appear
in a thermal model of the satellite with more nodes, are hardly necessary.

From the engineering standpoint, note that this satellite thermal model is
successful, insofar as it predicts that all nodal temperatures stay within
adequate ranges. In particular, nodes 9 and 10, corresponding to the boxes
with electronic equipment, stay within the range from $4$ to 23
$^{\circ}\mathrm{C}$.  These nodes are inner nodes with large thermal capacity
and, hence, are protected against the larger changes in the external heat
inputs. In contrast, the outer nodes are very exposed and undergo considerable
variation in temperature, with especially sharp changes at the beginning and
end of the eclipse.

\subsubsection{Second-order correction}

\psfrag{Temp}{{${T}_{(2)i}^\infty$ (K)}}
\begin{figure}
\begin{minipage}{8.4cm}
\centering{\includegraphics[width=8cm]{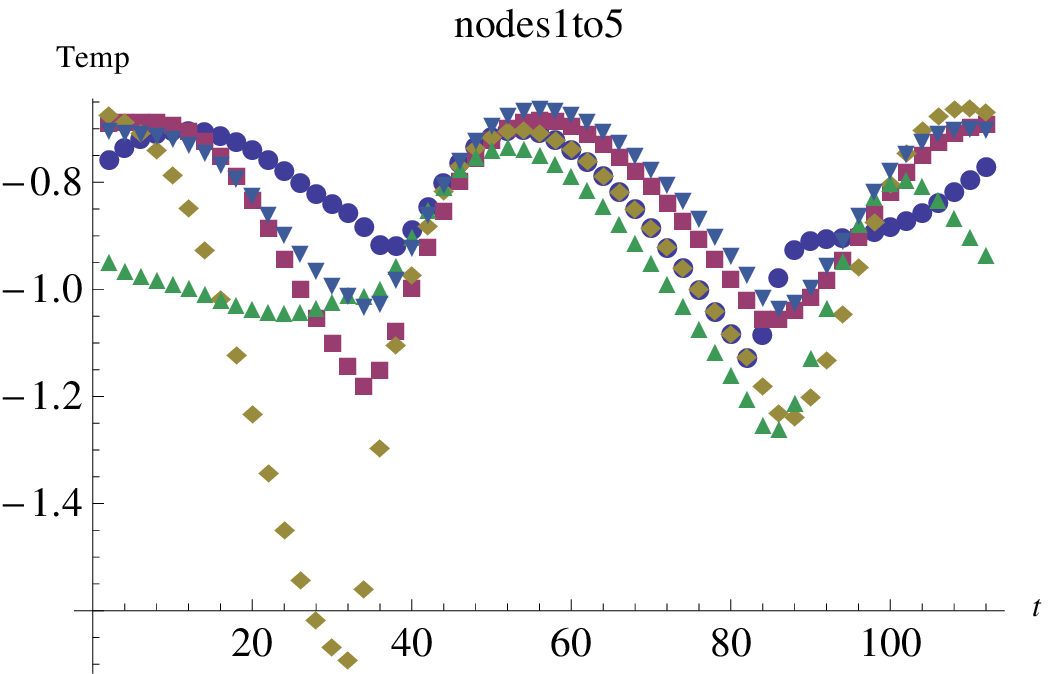}\\(a) nodes 1--5}
\end{minipage}
\begin{minipage}{8.4cm}
\centering{\includegraphics[width=8cm]{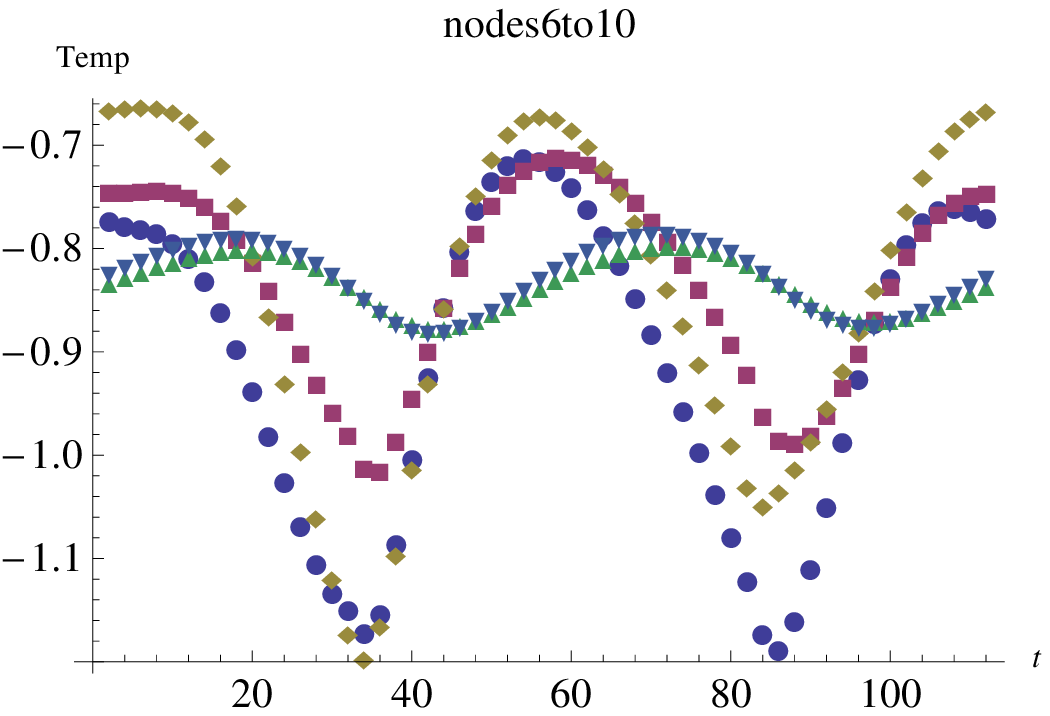}\\(b) nodes 6--10}
\end{minipage}
\\[2mm]
\caption{Second-order correction to the temperatures (same node order as in
  Fig.~\ref{Tevol}).}
\label{T2evol}
\end{figure}

According to Sect.~\ref{second-order}, the second-order perturbative
correction ${T}_{(2)}^\infty$ to the periodic stationary solution is obtained
by the same procedure as that for ${T}_{(1)}^\infty\,,$ but using a different
driving function $G$ that is computed from $\widetilde{T}$ and from
${T}^\infty_{(1)}$ itself.  The computations are straightforward and they
yield the correction plotted in Fig.~\ref{T2evol}.  This correction is
always negative, because the negative term in the expression for $G$,
Eq.~(\ref{tF}), dominates over the positive term.  The equation
(\ref{T_1_lim}) for ${T}_{(1)}^\infty$ and the corresponding equation for
${T}_{(2)}^\infty$ are both linear, so ${T}_{(1)}^\infty$ and
${T}_{(2)}^\infty$ are proportional to the respective driving functions; and
we can compare their magnitudes by comparing those driving functions, say,
comparing typical values of $\dot{Q}$ and $6 R_i\widetilde{T}_i^2
\,{T_{(1)i}^\infty}^2$.  This latter quantity can be roughly estimated as
$6\cdot 2\cdot 10^{-9} \cdot 300^2 \cdot 20^2$ W $\simeq 0.4$ W, whereas
$\dot{Q} \simeq 10$ W (Table~\ref{tab1}). Their ratio is about 25, which
roughly agrees with the ratio of ${T}_{(1)}^\infty$ to ${T}_{(2)}^\infty$, as
can be seen by comparing Fig.~\ref{Tevol} to Fig.~\ref{T2evol}.

The order of magnitude of the second-order correction suggests that higher
orders are not necessary, as we show next.

\subsection{Direct integration of the nonlinear equations}
\label{DNS}

Of course, the periodic solution $T^\infty(t)$ can also be obtained by direct
integration of the nonlinear Eqs.~(\ref{ODE}) with an adequate solver, based
on Runge-Kutta or other methods \cite{Kris3}. Since the nonlinear analysis
proves that the solution of Eqs.~(\ref{ODE}) will converge to the stationary
periodic solution \cite{NoDy1}, the numerical solver must be run until this
periodicity is established. Periodicity can be enforced by comparing the nodal
temperatures at the beginning and the end of every period and demanding that
they be equal, within some tolerance.  To do this, ESATAN$^\mathrm{TM}$
provides the routine SOLCYC \cite{ESATAN}.  We have employed this routine to
obtain the nonlinear equations' cyclic solution (which is established in 10
periods if the tolerance is set to 0.001).

\psfrag{Tdiff}{{$\D{T}_i$ (K)}}
\psfrag{nodes1and3and8}{{nodes 1, 3 and 8}}
\begin{figure}[!hb]
\centering{\includegraphics[width=8cm]{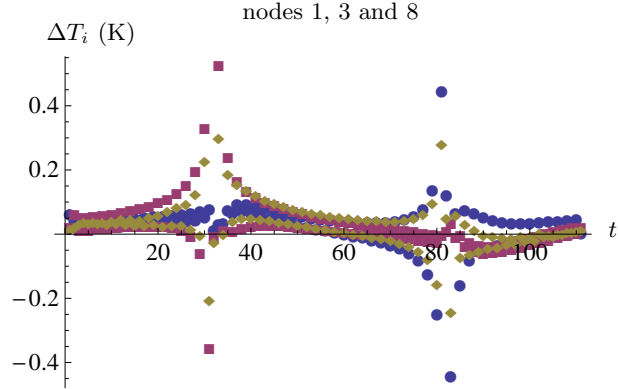}}
\caption{The error in the temperature computed up to the second order. Only
showing nodes 1 (dots), 3 (squares), and 8 (diamonds), which give rise to the
largest errors.}
\label{Terror}
\end{figure}

To compare the cyclic solution obtained by the linear method with the
``exact'' solution $T^\infty(t)$ obtained by SOLCYC, we quantify the deviation
by the vector of ``errors''
$$\D{T}(t) = T^\infty(t) - \left[\widetilde{T} + {T}_{(1)}^\infty(t) +
{T}_{(2)}^\infty(t)\right].$$ 
The largest components of $\D{T}$ are plotted in Fig.~\ref{Terror} (the
remaining components stay in the range $[-0.1,0.1]\,^{\circ}\mathrm{C}$).
There seem to be two branches for each node, but, in fact, it is an effect
produced by the high-frequency oscillations of each $\D{T}_i(t)$.  Notice that
the error is generally small when compared with ${T}_{(2)}^\infty(t)$; namely,
$|\D{T}(t)| \ll |{T}_{(2)}^\infty(t)|$ for each $t=k \T/n, \; k=0,\ldots,n-1$,
except near the two positions corresponding to the beginning and end of the
eclipse, where the heat inputs have discontinuities (Fig.~\ref{heat-in}).
These discontinuities induce oscillations of certain amplitude about the true
values of $T^\infty(t)$, due to the Gibbs phenomenon in the discrete Fourier
transform (Sect.~\ref{Fourier-anal}).  To suppress the oscillations, we would
need a specific method for treating the Gibbs phenomenon; for example, we
could use a smoother method for Fourier series summation, such as Ces\`aro
summation \cite{Fourier}

\section{Scalability and complexity of the linear method}
\label{scale}

The ten-node model studied in Sect.~\ref{ten-node} is too small to pose a
computational problem, whether we employ the linear method or directly
integrate nonlinear Eqs.~(\ref{ODE}).  To assess the practical applicability
of the linear method, we must study how it scales to realistic sizes and then
compare it with the direct integration of Eqs.~(\ref{ODE}).  Naturally, the
first step in the method is to compute the steady-state temperature
$\widetilde{T}_i\,,$ but this computation, arguably, is not a substantial part
of the whole process and we do not consider it (it may take, say, between a
few percent and one fith of the whole process, depending on the
circumstances).  The size of an orbiting spacecraft thermal model can be
scaled with respect to the number of nodes, $N$, or the number of different
positions in the orbit, $n$. However, while realistic models must have many
nodes, an $n$ of about one hundred can generally be suitable.  Note, in
particular, that the condition that $N \ll n$ in the ten-node model is likely
to be reversed in realistic models.  For these reasons, the scaling with
respect to $N$ is more relevant.

For the moment, let us neglect any special feature of Eqs.~(\ref{ODE}), such
as possible coefficient sparsity.  Then, the computational complexity of
numerically integrating those equations is of order $N^2 k$, where $k$ is the
number of time steps taken.  On the other hand, the complexity of the
numerical integration of linear Eqs.~(\ref{linODE-pert}) is also of order $N^2
k$ [excluding the computation of $J(t)$].  We can employ the explicit
integral, Eq.~(\ref{T_1}), which can be calculated with the trapezoidal rule,
for example, but this does not reduce the complexity of the computation.
However, for the stationary solution, Eq.~(\ref{T_1_lim}), its expression as an
integral over a period $\T$ sets the number of time steps to $n$, which is
advantageous if $n < k$, where $k$ now is the number of steps necessary for
some initial temperatures to relax to the stationary solution.  Since we have
to evaluate the integral in Eq.~(\ref{T_1_lim}) for several times $t$, it is
preferable to use the FFT, as discussed in Sect.~\ref{Fourier-anal}, so that
the total operation count is of order $N^2\, n \log n$.

However, we have not taken into account matrix operations of considerable
complexity. For example, the discrete Fourier transform,
Eq.~(\ref{d_Fourier-sol}), involves the {\em inverse} of an $N \times N$
matrix ($-J$ plus a multiple of the identity matrix), and the inversion of a
matrix is generally a process of order $N^3$ \cite{Num_rec}.  If we have to
employ a process of order $N^3$, we may as well diagonalize $J$, because the
diagonalization of a matrix also is generally a process of order $N^3$
\cite{Num_rec} and the diagonalization of $J$ has several uses.  Let us assume
that we carry out this diagonalization and determine the independent thermal
modes, which can then be employed to express Eq.~(\ref{T_1_lim}) as
Eq.~(\ref{T_1_eigenvec}) or to simplify the matrix operations in
Eq.~(\ref{d_Fourier-sol}).  If we roughly compare the computational complexity
of order $N^3$ with the numerical integration complexity of order $N^2 k$, we
deduce that the diagonalization is worthwhile when $N < k$.  Using the above
estimate $n \sim 100$ and taking as relaxation time $k \sim 5n$ (the rough
value for the ten-node model), we deduce that the determination of thermal
modes can be useful just as a computational procedure for models with a few
hundred nodes.

Let us now consider that $J$ is surely a sparse matrix, so iterative matrix
methods can take advantage of this characteristic.  In fact, there are two
degrees of sparsity in $J$, associated with conduction or radiation coupling
terms. The conductance matrix $K$ has to be very sparse, because conduction is
a local process, so each node can only be coupled to a few nodes. In contrast,
radiation is a nonlocal process and couples any pair of nodes that has a
nonvanishing view factor. In addition to the different sparsity of conduction
and radiation coupling matrices, there are two other circumstances that make
them different: (i) we are assuming that the conductive coupling matrix just
depends on material properties, so it is independent of the reference
temperatures $\widetilde{T}_i$; (ii) the conduction coupling terms are
significantly larger than the radiation coupling terms for the natural values
of those temperatures.  All of this suggests separating the conduction and
radiation parts of $J$, and then diagonalizing the conduction part.

Based on the study in Sect.~\ref{heur}, the diagonalization of the conduction
part of $J$ boils down to the diagonalization of the (generalized) Laplacian
matrix $-C^{-1/2}KC^{-1/2},$ and this matrix is sparse. Suitable iterative
algorithms to perform this diagonalization are, for example, the Lanczos
\cite{Gol-Van} or the Davidson \cite{David} algorithms.  These algorithms are
particularly useful when only a few of the largest or smallest eigenvalues are
needed. This is indeed our case, as only the slower modes are expected to
contribute to ${T}_{(1)}^\infty$ and ${T}_{(2)}^\infty\,.$ Once the conduction
part of $J$ has been diagonalized, in the sense that the lowest eigenvalues
and eigenvectors of the corresponding Laplacian matrix are known, the
radiation part of $J$ can be treated as a perturbation, using matrix
perturbation methods \cite{pert-methods,Gol-Van}.

Iterative matrix methods, combined with matrix perturbation methods or other
methods, if necessary, can reduce the order $N^3$ to $N^2$ or even to almost
linear and so allow us to diagonalize the Jacobians for the largest values of
$N$ that appear in current thermal spacecraft models.  Of course, the sparsity
of thermal coupling matrices also facilitates the direct integration of the
nonlinear Eqs.~(\ref{ODE}). However, their nonlinearity prevents one from
taking advantage of the above mentioned approximations methods, for example,
the reduction to the small subspace of slow modes, or the splitting into
conduction and radiation in which the latter is treated as a matrix
perturbation.  Moreover, the linearization is useful in various respects. For
example, the overall relaxation time, given by the eigenvalue of smallest
magnitude, can be effectively bounded by inequalities \cite{ineq} and some of
these bounds can be found with little computational effort.

\section{Summary and discussion}
\label{sec:4}

We have studied the evolution of the thermal state of an orbiting spacecraft
and developed a linear approach to this problem that is based on a rigorous
perturbative treatment of the exact nonlinear equations.  The first-order
perturbation equations, Eqs.~(\ref{linODE-pert}), constitute the basic linear
system, which can be applied to higher orders after calculating the
corresponding driving terms.  As the Jacobian matrix of the nonlinear
equations has negative eigenvalues, the linear equations describe the
relaxation to a stationary thermal state, namely, a periodic solution that is
independent of the initial conditions and only depends on the external heat
input.  This relaxation is similar to the relaxation to steady-state under
constant external heat load.

We have shown that the perturbative treatment reveals the scope of a common
linearization procedure of a heuristic nature, in which the nonlinear
equations are rendered linear by the definition of radiation conductances
(Sect.~\ref{heur}). If one previously calculates, with the correct nonlinear
Eqs.~(\ref{bODE-pert}), the reference steady-state condition that corresponds
to the average external heat input, the deviation from that steady-state is
well approximated by the linear equations with radiation conductances.  The
Jacobian matrix corresponding to radiation conductances, obtained in
Eqs.~(\ref{nJij}) and (\ref{nJii}), is related to a symmetric matrix and,
therefore, is easier to diagonalize.  Furthermore, this relation implies that
the thermal modes are normal, like the vibrational modes of a mechanical
system. Although the notion of radiation conductance is just an approximation,
it serves nonetheless to show that the Jacobian matrix is diagonalizable and
has real eigenvalues.

The diagonalization of the Jacobian matrix is useful for the computation of
the stationary thermal state and also provides information on the relaxation
to that state, because the relaxation times of the thermal modes are the
inverses of the eigenvalues.  These times span a considerable range, but the
longest times are much more significant than the shortest times, because the
latter depend on the details of the lumped-parameter thermal model employed
whereas the former are essentially independent of it.  In fact, a thermal
model that has more nodes and therefore more details also has more thermal
modes; but the slowest modes, which correspond to temperature changes in large
parts of the spacecraft, are hardly affected by the details, whereas the fast
modes can be significantly altered.  The slowest mode, in particular,
corresponds to a simultaneous but non-uniform increase (or decrease) of the
temperature throughout the spacecraft and is hardly altered by small-scale
changes.

The computation of the stationary thermal state with the linear method relies
on an explicit integral, Eq.~(\ref{T_1_lim}), or a Fourier expansion,
Eq.~(\ref{d_Fourier-sol}). Given a sampling of the thermal driving function at
equal time intervals, the periodic solution can be obtained through two
discrete Fourier transforms: a direct transform to get the Fourier
coefficients of the driving function and an inverse transform of the
coefficient vector multiplied by a suitable matrix
(Sect.~\ref{Fourier-anal}). Of course, the discrete Fourier transforms are
best performed with a FFT algorithm. This computation is more efficient than
the numerical computation of the integral, Eq.~(\ref{T_1_lim}), if we need the
values of the temperatures at all the given sampling times. However, the
Fourier transform presents the Gibbs phenomenon, associated with sudden
variations of the heat loads, as occur at eclipse times, for example. The
Gibbs phenomenon introduces errors, but these errors could be suppressed with
special methods.

The computation of the thermal modes and the stationary thermal state for a
satellite ten-node thermal model confirms the validity of the linear method
for a minimal but realistic model.  The relaxation times span a considerable
range, between 55 seconds and nearly one hundred minutes. Of course, the
latter time must be almost independent of the particular thermal model used,
whereas the former has no intrinsic significance, and, if the number of nodes
grew, that time would shrink (thus further expanding the range of relaxation
times).  The slowest mode corresponds to node temperature increments with the
same sign (positive by convention), whereas the increments corresponding to
other modes have both signs. The periodic variation in the external heat input
(Fig.~\ref{heat-in}) excites the thermal modes and produces a definite pattern
of stationary temperature oscillations, well approximated by the first-order
solution (Fig.~\ref{Tevol}). The second-order correction is small compared to
the first-order solution, but it is worth computing, as it reaches
1.7~K. Higher order corrections are essentially negligible, but the error due
to the Gibbs phenomenon at the eclipse positions reaches 0.6 K (at the most).

Focusing on the computational aspects of the linear approach, we have studied
how it scales with the number $N$ of nodes and the number $n$ of sampling
positions on the orbit.  If the Jacobian matrix is dense, the complexity of
the corresponding matrix operations is of order $N^3$. It is convenient to
employ just one matrix operation, namely, the diagonalization of the Jacobian
matrix, because then only a few of the slowest modes are needed for the
remaining operations, so these have negligible complexity.  The complexity of
a direct numerical integration of the nonlinear equations is of order $N^2 k$,
$k$ being the number of time steps necessary for relaxation.  For a
low-altitude orbit, $k$ is expected to be on the order of one thousand, as for
our Moon-orbiting satellite.  Therefore, the linear method would be
computationally effective as just an integration method only for models with a
few hundred nodes.  At any rate, the Jacobian matrix can be assumed to be
sparse, and its conduction part can be assumed to be especially sparse, in
addition to being the larger part of the Jacobian matrix and also being
independent of the orbit. As the orbit may be subjected to changes in the
planning of a mission, a convenient strategy probably is to diagonalize the
conduction part at the outset and, when needed, add the radiation part within
some approximation scheme.  This strategy can be far more efficient than
integrating the nonlinear equations each time.

Moreover, the strength of the linear approach lies with the insight that it
provides about the thermal behavior of the spacecraft, as embodied by the
decomposition of its thermal modes, of which only the slowest ones are
significant.  These significant modes can actually be obtained with a reduced
thermal model using few nodes.  Therefore, the linear approach is especially
useful in the context of reduced models. Furthermore, it provides a method for
model reduction based on the mode decomposition: this decomposition can be
used to group nodes. Indeed, there is a technique for graph partitioning based
on the eigenvalues and eigenvectors of the Laplacian matrix of the graph
\cite{graph_eigenvec,graph_part}. According to Sect.~\ref{heur}, this
technique is applicable to the Jacobian matrix, but the details of this
application are beyond the scope of the present paper and are left for future
work.

Finally, our linear approach can surely be applied to other cyclic heating
processes that involve radiation heat transfer.

\subsection*{Acknowledgments} 
We thank Isabel P\'erez-Grande for bringing Ref.~\citenum{MiPe} to our attention.


\begin{thebibliography}{99}
\section*{References}

\bibitem{Kreith}
F.~Kreith, 
\textit{Radiation Heat Transfer for Spacecraft and Solar Power Plant Design}.
Intnal.\ Textbook Co., Scranton, Penn. (1962)

\bibitem{therm-control}
C.A.\ Wingate, \textit{Spacecraft Thermal Control}. In: \textit{Fundamentals
of Space Systems}, V.L.~Pisacane and R.G.~Moore (eds.), Oxford Univ.\ Press
(1994)

\bibitem{therm-control_2}
D.G.~Gilmore (ed.), \textit{Spacecraft Thermal Control Handbook}.
The Aerospace Press, El Segundo (2002)

\bibitem{therm-control_3}
C.J.\ Savage, \textit{Thermal Control of Spacecraft}. In: \textit{Spacecraft
Systems Engineering}, Third Edition, P.~Fortescue, J.~Stark and G.~Swinerd
(eds.), Wiley, Chichester (2003)

\bibitem{anal-sat}
K. Oshima and Y. Oshima, {\em An analytical approach to the thermal design of
spacecrafts}. Rep.\ No.\ 419, Inst.\ of Space and Aeronautical Science of
Tokio (1968)

\bibitem{anal-sat_2}
J.-R. Tsai, 
\textit{Overview of satellite thermal analytical model}.
Journal of Spacecraft and Rockets, \textbf{41},
120--125 (2004)

\bibitem{IDR}
I.~P\'erez-Grande, A.~Sanz-Andr\'es, C.~Guerra and G.~Alonso,
\textit{Analytical study of the thermal behaviour 
and stability of a small satellite}.
Applied Thermal Engineering, \textbf{29}, 2567--2573 (2009)

\bibitem{NoDy}
J.~Gaite, A.~Sanz-Andr\'es and I.~P\'erez-Grande, 
\textit{Nonlinear analysis of a simple model of temperature evolution 
in a satellite}.
Nonlinear Dynamics, \textbf{58}, 405--415 (2009)

\bibitem{NoDy1}
J.~Gaite, 
\textit{Nonlinear analysis of spacecraft thermal models}.
Nonlinear Dynamics, \textbf{65}, 283--300 (2011) 

\bibitem{Kris1}
C.K. Krishnaprakas, 
\textit{Application of accelerated iterative methods for solution of thermal
models of spacecraft}.
Journal of Spacecraft and Rockets, \textbf{32},
608--611 (1995)

\bibitem{Kris2}
C.K. Krishnaprakas, 
\textit{Efficient solution of spacecraft thermal models using preconditioned
conjugate gradient methods}. 
Journal of Spacecraft and Rockets, \textbf{35},
760--764 (1998)

\bibitem{MiPe}
M. Milman and W. Petrick,
\textit{A note on the solution of a common thermal network problem encountered
in heat-transfer analysis of spacecraft}.
Applied Mathematical Modelling, \textbf{24}, 861--879 (2000)

\bibitem{Kris3}
C.K. Krishnaprakas, 
\textit{A comparison of ODE solution methods for spacecraft thermal
problems}. 
Heat Transfer Engineering, \textbf{19}, 103--109 (1998)

\bibitem{ESATAN} \textit{ESATAN-TMS Thermal Engineering Manual} and 
\textit{User Manual}. Prepared by
ITP Engines UK Ltd., Whetstone, Leicester, UK (2009)

\bibitem{Ber-Plem} A.\ Berman and R.J.\ Plemmons, \textit{Nonnegative Matrices
in the Mathematical Sciences}. Classics in Applied Mathematics, vol.~9, SIAM
(1994)

\bibitem{pert-methods}
E.J. Hinch, \textit{Perturbation methods}, Cambridge Texts in Applied
Mathematics (1991)

\bibitem{Chung}
F.R.K. Chung, \textit{Spectral Graph Theory}. Providence, RI:
Amer. Math. Soc. (1997)

\bibitem{graph_eigenvec} 
T. Biyikoglu, J. Leydold and P.F. Stadler, 
\textit{Laplacian Eigenvectors of Graphs: Perron-Frobenius and Faber-Krahn
Type Theorems}. Lecture Notes in Mathematics 1915, Springer (2007)

\bibitem{Fourier} 
H.F. Davis, \textit{Fourier Series and Orthogonal Functions}. Dover
Publications (1989) 

\bibitem{Num_rec} 
W.H. Press, S.A. Teukolsky, W.T. Vetterling, and B.P. Flannery, 
\textit{Numerical Recipes: The Art of Scientific Computing}.
Cambridge University Press, 3rd edition (2007)

\bibitem{Gol-Van} 
G.H. Golub and C.F. Van Loan, \textit{Matrix Computations}.
The Johns Hopkins U. Press, Baltimore (1996)

\bibitem{David}
M. Crouzeix, B. Philippe and M. Sadkane, \textit{The Davidson Method}. SIAM
Journal on Scientific Computing, \textbf{15}, 62--76 (1994)

\bibitem{ineq} 
G.-X. Tian and T.-Z. Huang, 
\textit{Inequalities for the minimum eigenvalue of M-matrices}.
Electronic Journal of Linear Algebra, \textbf{28}, 291--302 (2010)

\bibitem{graph_part} 
U. Luxburg, \textit{A tutorial on spectral clustering}. Statistics and
Computing, \textbf{17}, 395--416 (2007)

\end{thebibliography}
\end{document}